\documentclass[fleqn,usenatbib]{mnras}

\usepackage[T1]{fontenc}
\usepackage{ae,aecompl}
\usepackage{multirow}
\usepackage{dirtytalk}

\usepackage{graphicx}	
\usepackage{amsmath}	
\usepackage{amssymb}	
\usepackage{pdfpages}   

\graphicspath{{figures/}}

\newcommand{\dd}{\text{d}}
\newcommand{\bfrac}[2]{\left( \frac{#1}{#2} \right)}
\newcommand{\ufrac}[2]{\bfrac{#1}{\rm #2}}
\newcommand{\avg}[1]{\langle #1 \rangle}

\newcommand{\new}[1]{#1}
\newcommand{\stefan}[1]{\textcolor{magenta}{[Stefan: #1]}}

\newcommand{\primordial}{{\it primordial }}
\newcommand{\astrophysical}{{\it astrophysical }}
\newcommand{\Uniform}{{\it Uniform }}
\newcommand{\stardensity}{{\it star density }}
\newcommand{\heesen}{{\it dwarf }}
\newcommand{\uniform}{{\it uniform }}
\newcommand{\wind}{{\it wind }}
\newcommand{\windSNR}{{\it wind+SNR }}
\newcommand{\MW}{{\it NE2001 + JF12 }}

\newcommand{\unitDM}{{\rm\ pc \ cm^{-3}}}
\newcommand{\unitRM}{{\rm\ rad \ m^{-2}}}

\newcommand{\nISM}{ n_{\rm ISM}}
\newcommand{\RMobs}{\text{RM}_{\rm obs}}
\newcommand{\RMEG}{\text{RM}_{\rm EG}}
\newcommand{\RMMW}{\text{RM}_{\rm MW}}
\newcommand{\RMIGM}{\text{RM}_{\rm IGM}}
\newcommand{\RMProg}{\text{RM}_{\rm prog}}
\newcommand{\RMHost}{\text{RM}_{\rm host}}
\newcommand{\DMobs}{\text{DM}_{\rm obs}}
\newcommand{\DMEG}{\text{DM}_{\rm EG}}
\newcommand{\DMMW}{\text{DM}_{\rm MW}}
\newcommand{\DMIGM}{\text{DM}_{\rm IGM}}
\newcommand{\DMProg}{\text{DM}_{\rm prog}}
\newcommand{\DMHost}{\text{DM}_{\rm host}}

\newcommand{\ana}{A\&A}
\newcommand{\PASP}{PASP}
\newcommand{\aas}{A\&AS}
\newcommand{\aarv}{A\&ARv}

\title[FRBs \& IGMFs]{Fast Radio Burst dispersion measures and rotation measures and the origin of intergalactic magnetic fields}

\author[Hackstein et al.]{
{S. Hackstein$^{1}$}\thanks{E-mail: stefan.hackstein@hs.uni-hamburg.de},
{M. Br\"uggen$^{1}$},
{F. Vazza$^{1,2}$},
{B. M. Gaensler$^{3,4}$},
{V. Heesen$^1$}
\\
$^{1}$Hamburger Sternwarte, University of Hamburg, Gojenbergsweg 112, 21029, Germany\\
$^{2}$University of Bologna,  Department of Physics and  Astronomy,
Via Gobetti 93/2, I-40129, Bologna, Italy;\\$^{\phantom{22}}$Istituto di Radioastronomia, INAF, Via Gobetti 101,40129 Bologna, Italy\\
$^3$Dunlap Institute for Astronomy and Astrophysics, University of Toronto, Toronto, ON M5S 3H4, Canada \\
$^4$Department of Astronomy and Astrophysics, University of Toronto, Toronto, ON M5S 3H4, Canada
}
\date{Accepted 2019 July 12. Received 2019 July 10; in original form 2019 April 18}

 \hypersetup{draft}

\begin{document}
\label{firstpage}
\pagerange{\pageref{firstpage}--\pageref{lastpage}}
\maketitle

\begin{abstract}
We investigate the possibility of measuring intergalactic magnetic fields using the dispersion measures and rotation measures of fast radio bursts.
With Bayesian methods, we produce probability density functions for values of these measures.
We distinguish between contributions from the intergalactic medium, the host galaxy and the local environment of the progenitor. 
To this end, we use constrained, magnetohydrodynamic simulations of the local Universe to compute lines-of-sight integrals from the position of the Milky Way. 
In particular, we differentiate between \new{predominantly} astrophysical and primordial origins of magnetic fields in the intergalactic medium.
We test different possible types of host galaxies and probe different distribution functions of fast radio burst progenitor locations inside the host galaxy. 
Under the assumption that fast radio bursts are produced by magnetars, we use analytic predictions to account for the contribution of the local environment.
We find that less than 100 fast radio bursts from magnetars in stellar-wind environments hosted by starburst dwarf galaxies at redshift $z \gtrsim 0.5$ suffice to discriminate between \new{predominantly} primordial and astrophysical origins of intergalactic magnetic fields.
However, this requires the contribution of the Milky Way to be removed with a precision of $\approx 1 \unitRM$.
\new{We show the potential existence of a subset of fast radio bursts whose rotation measure carry information on the strength of the intergalactic magnetic field and its origins.}
\end{abstract}





\section{Introduction}

Fast radio bursts (FRBs) are impulsive bursts in the radio sky of very short duration (0.1 - 10 ms) with frequencies of about 1 GHz, observed down to 400 MHz \citep{lorimer2007}.
Their observed dispersion measure ($\text{DM}$) exceeds the contribution of the Milky Way (MW), implying an extragalactic origin.
Their short duration suggests an emitting region of the order of 100 km, suggesting a neutron star origin. 
Such a small region only allows for small intrinsic variation of e. g. the polarization angle, used to observe the Faraday rotation measure ($\text{RM}$), which is directly related to the magnetic field strength along the line of sight (LoS).
FRBs are hence potential probes for the intergalactic medium (IGM) and interstellar medium in the MW and in the host galaxy, especially in the local environment of the FRB progenitor \citep[see e. g.][]{zheng2014,ravi2016,keane2016host}.
In this work, we investigate whether FRBs with observed $\text{RM}$s can be used to derive information on the origin of intergalactic magnetic fields (IGMFs).\\

Currently, the most widely accepted constraints on the comoving strength of magnetic fields in voids stem from observations of the CMB \citep
[$B \lesssim 4.4 \cdot 10^{-9} \rm\ G$,][]{Planck2015PMF} and of TeV-Blazars \citep
[$B \gtrsim 3 \cdot 10^{-16} \rm\ G$,][]{neronov2010}, about seven orders of magnitude apart.
For a summary of constraints on the magnetic field strength and coherence lengths, see \citet{taylor2011} or \citet{dzhatdoev2018}.\\

A number of processes have been proposed to generate cosmic magnetic fields.
Primordial scenarios consider processes in the early Universe, mostly prior to the re-combination epoch, e. g. during phase transitions or inflation \citep[e.g.][]{2009IJMPD..18.1395C,2010PhRvD..82h3005K,Subramanian2016}. 
Another possible scenario is the generation of magnetic fields during early galaxy formation, e. g. by feedback of active galactic nuclei \citep[e. g.][]{vazza2017}  or winds from compact galaxies \citep{Kronberg1999,donn09,dubois2010}.
For a detailed overview on the different models, see e. g. \citet{widrow2002}.
These two scenarios result in severely different predictions for the magnetic field strengths in voids of the large-scale structure. 
In reality, it is likely that both scenarios contribute to the origin of cosmic magnetic fields.
Measuring their strength would allow to put reasonable constraints on the origin of those fields \citep[e.g.][]{vazza2017}. \\

Since their first discovery \citep{lorimer2007}, there has been a large number of studies addressing the nature and origin of FRBs \citep[e. g.][]{zhang2013,ravi2018,marcote2019}, see  \citet{katz2016-review}, \citet{Lorimer2018} and \citet{Petroff2019review} for reviews. 
\citet{ravi2019} have summarized the expected progress in the coming decade.

So far, two repeating sources have been identified \citep{spitler2016,scholz2016,Amiri2019} that rule out cataclysmic scenarios, at least for those events. 
Many more discoveries are expected in the very near future.
Repeating signals allow to test time dependence of their properties, making them the subject of intensive studies \citep[e. g.][]{lu2018,hessels2018,lyutikov2019,li2019,houben2019,Yang2019}.

Still, very little is known about the population and origin of FRBs \citep[e. g.][]{Katz2018,caleb2018,palaniswamy2018,caleb2019, James2019}.
To keep track of all the different models, they are collected in the living theory catalogue of FRBs \citep{platts2018livingcat}.
Here, we investigate the application of FRBs as probes of cosmic magnetism, with a few priors on their possible origin. 
We present a framework that can be used to compare observations to theory to make quantitative inferences.\\

At this point, only a few FRBs have observed $\text{RM}$s.
Once the next generation of telescopes, such as e. g. CHIME/FRB, FAST, MeerKat and SKA, begin their surveys, this number is expected to increase drastically \citep{Jonas2009,Nan2011,Keane2013,Macquart2015,CHIME/FRB2018}.\\

\citet{Akahori2016} produced predictions for the intergalactic $\DMIGM$ and $\RMIGM$ of FRBs from the IGM
They use numerical simulations for the large-scale structure and the IGMF to test whether combining both measurements yields information on IGMFs.
Their results show that the $\text{RM}$ is dominated by the hot gas in clusters while the dominant contributor to $\text{DM}$ changes with redshift.
Still, they show that the radial component of the density-weighted IGMF strength in filaments can be inferred from combined measurements within a factor of $\sim 2$.

\citet{Vazza2018_FRB} investigate the variance in $\RMIGM$ due to the assumed magneto-genesis model.
They assume astrophysical or primordial origin of cosmic magnetic fields, similar to the models used in this work.
For FRBs located at a redshift of $z=1$, they find differences in $\langle \RMIGM \rangle$ between the models of about one order of magnitude. 
In principle, this allows to draw conclusions on the strength of the IGMF.
However, it is unclear at which redshift the observed signal reveals most information. \\

\citet{Walker2018}  provide   a   framework similar   to the  one  presented  in  this  paper.  
They obtain  predictions of $\text{DM}$ in  form  of  likelihood  functions for the different contributing regions.
They use these to derive estimates on the redshift of FRBs, which mostly agree with $z\approx 0$ in the lower bounds.
Thus they conclude that the observed $\DMobs$ can only be used to infer an upper limit on the redshift.
This is in agreement with several other studies on that topic \citep{dolag2015,Niino2018,luo2018,Pol2019}.
We note that the framework presented here easily allows one to infer the redshift of an FRB from its $\text{DM}$, similar to the findings of \citet{Walker2018} and \citet{Pol2019}.
However, our results are subject to the same uncertainties and cannot improve on previous results. \\

\begin{figure}
    \centering
    \includegraphics[width=0.48\textwidth]{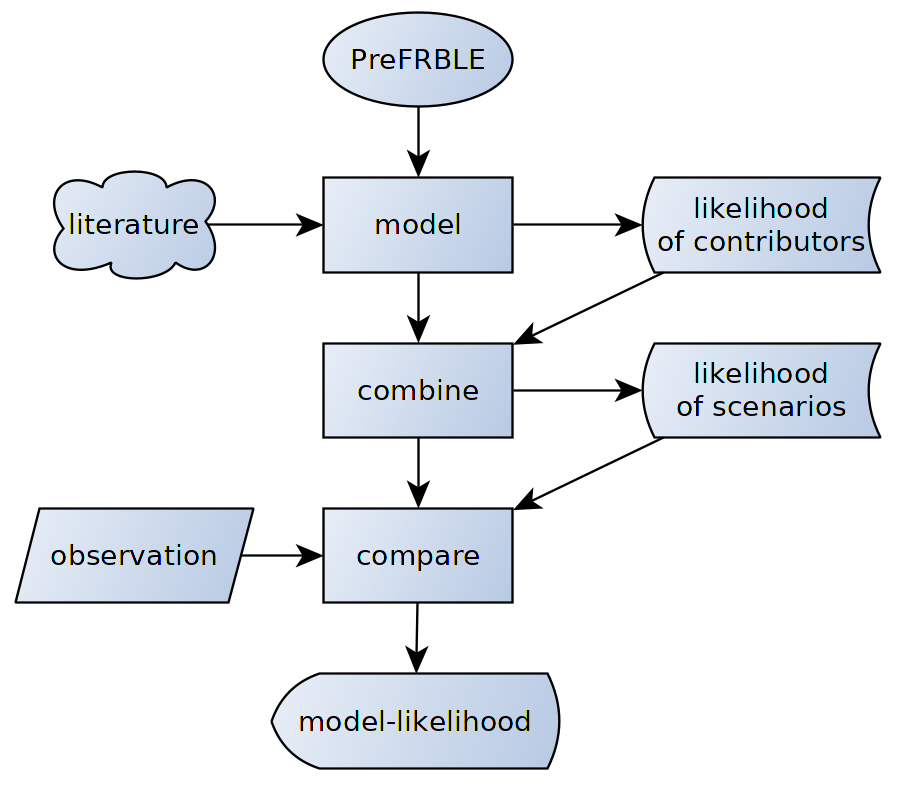}
    \caption{Flowchart to depict the basic structure of the inference presented in this work.
    We use results in the literature to model the contributions to $\text{DM}$ and $\text{RM}$ from different regions along the LoS of FRBs.
    These results are combined to predict the full measures expected at Earth in different scenarios for combination of contributor models.
    Finally, the results are compared to observational data to quantify and compare the posterior likelihood of several scenarios to produce the observed data.}
    \label{fig:flowchart}
\end{figure}

In this work, we show how to combine predictions of $\text{DM}$ and $\text{RM}$ for different regions along the LoS of FRBs and to compare them to the observed $\DMobs$ and $\RMobs$ in order to study the IGMF.
Fig.~\ref{fig:flowchart} shows an overview of the basic structure of the inference.

We improve on previous studies by use of constrained simulations of the local Universe that resemble different scenarios of magneto-genesis.
Further, by comparing the individual contributions to DM and RM along the LoS, considering redshifts out to $z=6$.\\

This paper is organized as follows. 
In Sec.~\ref{sec:models} we explain how we model the different contributions to $\text{DM}$ and $\text{RM}$ along the LoS of FRBs and how to compute their likelihood functions.
In Sec.~\ref{sec:model_results} we discuss the results of the individual models of the contributing regions.
In Sec.~\ref{sec:combine} we combine the predictions of all contributors to predict observed $\DMobs$ and $\RMobs$. 
We show how these can be used to interpret $\DMobs$ and $\RMobs$ regarding the origin of IGMFs.
Finally, in Sec.~\ref{sec:discussion} we  discuss our results and in Sec.~\ref{sec:conclusion} we conclude.

\section{Models}
\label{sec:models}
\begin{table*}
\centering
    \begin{tabular}{c|c}
         mnemonic & physics  \\ \hline \hline
         \multicolumn{2}{l}{ \bf IGM: } \\ \hline
         \primordial & 3D-MHD model of the local Universe with strong uniform initial magnetic field of $B_0 = 1 \rm\ nG$ comoving \\
         \astrophysical & 3D-MHD model of the local Universe with very weak initial magnetic field and magnetic feedback of AGN \\ \hline
         \multicolumn{2}{l}{\bf Host Galaxy: } \\ \hline
         \Uniform & MW-like spiral galaxy model (NE2001 
         \& JF12) ,
         $n_{\rm FRB} = const.$ \\
         \stardensity & MW-like spiral galaxy model (NE2001 \& JF12),
         $n_{\rm FRB} \propto n_\star$ \\ 
         \heesen & starburst dwarf galaxy similar to IC10 or Host of FRB121102,
         $n_{\rm FRB} \propto n_\star$ \\ \hline
         \multicolumn{2}{l}{\bf Local Environment of Progenitor:} \\ \hline
         \uniform & magnetar in uniform ISM environment \\
         \wind & magnetar in environment dominated by stellar winds of seed star \\
         \windSNR & \wind plus contributions of SNR \\ \hline
         \multicolumn{2}{l}{\bf Milky Way: } \\ \hline
         \MW & best-fitting model for Galactic $\text{RM}$ (NE2001 \& JF12)
    \end{tabular}
    \caption{Summary of all models investigated in this work.
    $n_{\rm FRB}$ is the assumed number density of possible progenitor positions,
    $n_\star$ the number density of stars in the MW.
    NE2001 stands for the density model of thermal electrons in the MW presented in \citet{NE2001}.
    JF12 stands for the magnetic field model of the MW developed by \citet{jf12}.
    }
    \label{tab:models}
\end{table*}
In this section we describe the models investigated in this work and how we obtain the likelihood functions.
A summary of all models can be found in Tab. \ref{tab:models}.

\subsection{Intergalactic Medium}
\label{sec:IGM}
\subsubsection{Model}
\begin{figure}
    \centering
    \includegraphics[width=0.47\textwidth]{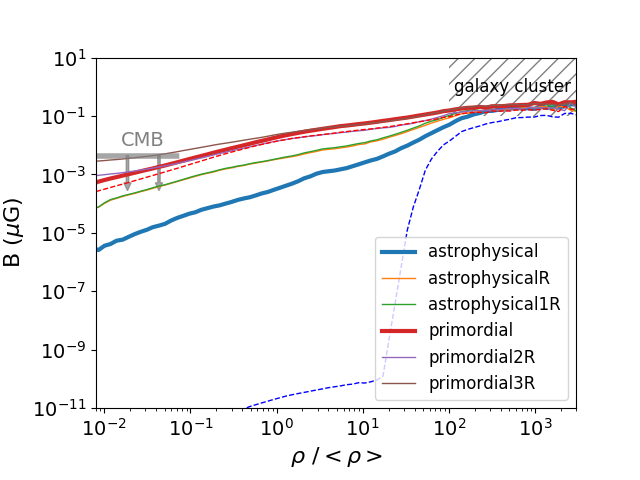}
    \caption{Relation between average magnetic field strength $B$ and gas density $\rho$ in the different IGM models investigated in \citet{Hackstein2018}.
    The two models used in this paper are drawn with thick lines.
    The dashed lines show results for the median of $B$, instead of the mean.
    We indicate the range of magnetic fields in clusters \citep[e. g.][]{Feretti2012} as well as the upper limit of fields in voids according to CMB observations by PLANCK, $B_0 \lesssim 4.4 \rm\ nG$ \citep{Planck2015PMF}.
    }

    \label{fig:B-rho}
\end{figure}
We model the IGM using constrained magnetohydrodynamic (MHD) simulations of the local Universe, produced with the ENZO code \citep{Bryan2014}.
The simulations start from initial conditions obtained from a procedure summarized by \citet{Sorce2016}.
The constraints applied in order to reproduce the local Universe at $z=0$ are fully described by \citet{Tully2013}.
Simulations have been produced within the Planck cosmology framework \citep[$\Omega_m = 0.307$, $\Omega_\Lambda = 0.693$, $h=0.677$, $\sigma_s = 0.829$,][]{Planck2014}.
Further information on the models can be found in \citet{Hackstein2018}, where they have been investigated in the context of propagation of cosmic rays. 
The three-dimensional datasets at $z=0$ are also publicly available at \href{https://crpropa.desy.de/}{https://crpropa.desy.de/} under 'Additional resources'. \\

We consider two different scenarios for the predominant genesis of IGMFs, \primordial 
vs. \astrophysical. 
To do so, we make use of the result of a single simulation, which considers the \primordial origin of IGMFs.
From that and from the \astrophysical model presented in \citet{Hackstein2018}, we extract the $|B| \propto\rho$ relation in Fig. \ref{fig:B-rho} \citep[cf. ][]{vazza2017}.
\new{The difference in $|B|(\rho)$ between the two models is most prominent at very low density, far away from the central cluster regions, where most active galactic nuclei reside.
    However, the contribution from these regions to the RM is likely far below the ionospheric foreground $\approx 1 \unitRM$, hence not observable.
    The most interesting regions are the vicinity of clusters, filaments and other regions, where $1 < \rho/\avg{\rho} < 200$. }

\new{The \primordial model starts with a uniform magnetic field 
    with comoving magnetic field strength $B_0 = 1$ nG, slightly below upper limits of the PLANCK collaboration, $B_0 \lesssim 4.4 \rm\ nG$ \citep{Planck2015PMF}.
    Note that \citet{Trivedi2014} derived an upper limit of $B_0 \lesssim 0.6 \rm\ nG$ using the CMB Trispectrum. 
    The \astrophysical model is initialized with a $B_0 = 10^{-8}\rm~nG$.
    Though this is below the lower limits obtained for present fields in voids, the final result of the simulation agrees with that limit, $B_{\rm void} \gtrsim 3\cdot 10^{-7} \rm~nG$ \citep{neronov2010}.
    In order to obtain magnetic fields that agree with observations of clusters, this model allows for additional seeding of magnetic fields by feedback of active galactic nuclei below redshift 4.
In order to obtain results for the \astrophysical model from the data of \primordial, we apply the ratio of average $|B|(\rho)$ for these two models as} correction factor on the magnetic field outside of galaxy clusters, where cosmic gas density $\rho < 200 \avg{\rho}$.
This allows us to test different prescriptions for three-dimensional magnetic fields in our cosmological volume with a limited consumption of computing time. 
\new{However, in this work we investigate only two models at the extreme ends of possible strengths of the IGMF in order to see whether FRBs carry any information on the IGMF.} 

\new{Note that the average of $|B|$ is dominated by the high values in a density bin.
	The median, shown as dashed lines in Fig. \ref{fig:B-rho}, reflects much better the huge difference in the magnetic field outside high density structures.
	Using the median ignores possible high values of $|B|$ within a density bin, hence understimates the magnetic field and the RM.
	The average instead is dominated by these high values and forces the magnetic field to values of similar strength, everywhere within the density bin.
	In this case, the results for the \astrophysical model are much closer to the \primordial, representing the most pessimistic case to tell the two extreme models apart.
	Hence, the use of the average instead of the median strengthens the conclusion, that observation of FRBs can be used to distinguish between these two models.
}
\\

We note however, that the \primordial model we probe here is initialized with a uniform field, whose topology is preserved in low-density regions. 
This can affect the distribution of $\text{RM}$ from FRBs in the local Universe, making smaller values less probable.
This is, because the contributions from different parts of the IGM are less likely to cancel out each other.
In App. \ref{app:uniform} we investigate how that influences the final results and find a negligible impact on observable $\RMobs \gtrsim 1 \unitRM$ \citep[see also][]{Vazza2018_FRB}. \\

The use of numerical simulations will improve our results over those of \citet{Walker2018} and \citet{Niino2018}
by accounting for the uncertainty that arises due to inhomogeneities in the IGM.
Like \citet{Akahori2016}, we apply the usual method of cosmological data stacking \citep[e.g.][]{daSilva2000}. We reconstruct the cosmic space from redshift $z=0$ to $z=6$ with use of simulation outputs at redshifts $z$ = 0.0, 0.2, 0.5, 1.5, 2.0, 3.0, and 6.0.
The LoS starts at redshift $z=6$ and traverses the simulated volume in a randomly oriented rectilinear path.
When the LoS reaches the corresponding redshift, the trajectory is continued in the next snapshot.
The final snapshot at $z=0$ is used from half the cosmic time towards the previous snapshot at $z=0.2$.
Finally, all values are corrected for a smooth evolution with redshift. \\

The $\DMIGM$ for a LoS with cosmological distance is
\begin{equation}
    \DMIGM = \int\limits_0^{d_{\rm FRB}} (1+z)^{-1} \ufrac{n_e}{cm^{-3}}\ufrac{dl}{pc} \rm pc\ cm^{-3} ,
\end{equation}
where $d_{\rm FRB}$ is the comoving distance to the FRB source and $n_e$ is the proper thermal electron density \citep{Xu2015}.
$\text{DM}$ measures the extra travel time of radiation at low frequencies due to dispersive effects in plasma.
Hence, it scales with redshift as $\text{DM} \propto (1+z)^{-1}$.

The $\RMIGM$ for a LoS with cosmological distance is
\begin{equation}
    \RMIGM = \frac{\Delta \Phi}{\Delta \lambda^2} 
    \approx
    0.81 \int\limits^0_{d_{\rm FRB}} (1+z)^{-2} \ufrac{B_\parallel}{\mu G} \ufrac{n_e}{cm^{-3}} \ufrac{dl}{pc} \rm rad\ m^{-2} ,
\end{equation}
where $B_\parallel$ the proper magnetic field component parallel to the LoS \citep{xu2014}.
$\text{RM}$ is the ratio of relative rotation of polarization angles $\Phi$ at different frequencies divided by the difference of the squared wavelength $\lambda$.
The former is not affected by cosmic expansion, therefore $\text{RM}$ scales with redshift as $\text{RM} \propto (1+z)^{-2}$. \\

The free electron density, $n_e$, is computed assuming full ionization and a mean molecular weight $\mu_e \approx 1.16$ of an electron for cosmic fractions of hydrogen and helium.

\subsubsection{Likelihood function}
We obtain the likelihood function of the IGM contribution from LoS integrals.
These are produced using the {\sc LightRay} function of the {\sc trident} package \citep{trident}.
This function extracts field values from data cells intersected by a LoS, defined by start and end positions in the three-dimensional volume.
It also computes the redshift that reflects the distance to the observer.
This way, it allows us to compute the redshift-corrected values along the LoS that contribute to the $\text{DM}$ and $\text{RM}$.
\\
These LoS start from the position of the MW, defined as the centre of the box in our constrained simulation of the local Universe.
They progress in evenly distributed directions defined by the {\sc HEALPIX} \citep{Gorski2005} tessellation of the sphere \citep[similar to][]{Stasyszyn2010}.
We use a total of 49152 LoS, corresponding to a pixel radius of $1-2^\circ$.
This allows us to resolve local structures while computation costs are kept at a minimum.
The total computation took about 1200 hours of CPU time. Differences in the likelihoo function are $< 0.1$ per cent compared to the next smaller tesselation of the sky with 12288 LoS.
Hence, the likelihood function is reasonably converged.
\\

The total path length of the LoS exceeds the size of the high-resolution portion of our constrained Universe, which is the central $(250 \rm\ Mpc)^3$ volume of the simulation.
Therefore, direction-dependent results beyond the first crossing of this region would be misleading.
Instead, for results at higher redshift ($z\gtrsim 0.1$) we investigate a statistical sample of LoS with random orientation.
These are obtained by stacking segments between random points taken from the constrained regions, until the LoS reaches the redshift of the current snapshot.
The process continues with the next snapshot, until the full LoS is built.
The final snapshot of the simulation is at $z=0$ and would not be used in the procedure described above.
Hence, it is used until half of the cosmic time towards the next snapshot at $z=0.2$, where $z \approx 0.9$. 
The fact that $\nabla \cdot \vec{B}$ is not conserved to 0 at the interfaces where we combine different segments of the LoS does not pose any problem for our analysis, as $< 1$ per cent of cells is affected by this problem. \\

The likelihood function is proportional to the amount of LoSs that deliver the same value.
Assuming an isotropic distribution of FRBs in the sky, the calculation is straightforward:
\begin{equation}
p(\text{DM}'|z) = \frac{ \oint \delta(\text{DM}(\theta,\phi;z) = \text{DM}') \dd\theta\dd\phi }{ \oint \dd\theta\dd\phi  }  \approx \frac{ N_{\rm \text{DM}'} }{ N_{\rm tot} } ,
\label{eq:likelihood_IGM}
\end{equation}
where $N_{\rm \text{DM}'}$ is the number of LoSs from a redshift $z$ with $\text{DM} \approx \text{DM}'$, $N_{\rm tot}$ is the total number of LoSs from that redshift.
The same holds for the $\text{RM}$.

\subsection{Host galaxy}
\label{sec:galaxy}
\subsubsection{Model}
To highlight the influence of host galaxies, we investigate two different types of galaxies, a spiral galaxy similar to the MW and a starburst dwarf galaxy similar to the host of FRB121102.
We note that this small number of models does not suffice to reflect the variety of different galaxies that are likely to host FRBs, but gives a rough estimate on the range of possible contributions. \\

Integrating the galaxy stellar mass function \citep{baldry2012} yields that 68 per cent of stars reside in galaxies of $10^{11} - 10^{12} M_\odot$, similar to the MW. 
Such galaxies are likely hosts, if FRBs are produced by magnetars \citep[e.g.][]{Popov2010,Katz2016, Metzger2017,Beloborodov2017,Metzger2019}.
We obtain predictions for the spiral host galaxy with use of the NE2001 model \citep{NE2001} for the thermal electron density, combined with the JF12 model \citep{jf12} for the magnetic field, where we use the best-fitting parameters for the MW.
\citet{luo2018} compared the results of NE2001 with the model of \citet{yao2017} and found that the overall statistics are rather similar, NE2001 tending to slightly higher values of $\text{DM}$.
Here we exclusively use the NE2001 model, which was also assumed by \citet{jf12}.

Though it has been argued that the NE2001 model is not good enough to exactly reconstruct the $\text{DM}$ foreground of pulsar data \citep[see][]{Xu2015}, it is a reasonable choice to obtain a decent statistical estimate.
Calculations have been performed using the {\sc Hammurabi} code \citep{hammurabi}, which computes the LoS integrals in evenly distributed directions on the whole sky seen from a given position to the edge of the galaxy model.

The likelihood for of a given value of $\DMHost$ and $\RMHost$ from the host highly depends on the position of the progenitor within the host, which is uncertain.
To account for that, a reasonable choice is to sample different possible positions and combine their predictions.
The positions are drawn randomly, following a probability density that we assume to be either uniform or proportional to the star density.
In particular, for the latter we use the combination of a thick and a thin disc of radius $R_i$ and scale height $Z_i$ with exponentially falling star density 
\begin{equation}
    n_{\rm star}(R,Z) \propto \exp\left( -\frac{R}{R_i} - \frac{|Z|}{Z_i} \right),
\end{equation}
using the best fit parameters obtained for distribution of M dwarfs in the MW, i. e. $R_{\rm thick} = 3.6$ kpc, $Z_{\rm thick} = 0.9$ kpc, $R_{\rm thin} = 2.6$ kpc and $Z_{\rm thin} = 0.3$ kpc \citep{juric2008}. \\

Dwarf irregular galaxies in a starburst phase, which we will refer to as starburst dwarf galaxies thereafter, have high star-formation rates, hence their stellar population is relatively young.
Magnetars have short lifespans, $\approx 10^{4} \rm\ yr$ \citep[e. g.][]{Beniamini2019}, and are produced by massive stars, $20-45 ~\rm M_{\sun}$ \citep{chabrier2003} that have rather short lifetimes, $\sim 10^7 \rm\ yr$ \citep[e. g.][]{wit2005}.
This makes starburst dwarf galaxies a likely host for FRBs produced by magnetars.

The first localized FRB121102 was indeed found to reside in such a starburst dwarf galaxy, having a high star-formation rate, low metallicity and no active  galactic nucleus \citep{chatterjee2017,tendulkar2017host}
Low-mass and low-metallicity galaxies with high star-formation rates were also found to be over-represented hosts of gamma-ray bursts and superluminous supernovae at low redshift \citep[e. g.][]{fruchter2006,vergani2015,perley2016}.

A well-studied starburst dwarf galaxy in the local Universe is IC~10, which is at $0.8$~Mpc distance.
It is the only member of the Local Group that is currently in the starburst phase. 
Its properties are very similar to that of the host of FRB121102 \citep[e. g.][]{leroy2006IC10,richer2001ic10,magrini2009ic10}. 
We  use of the magnetic field model of \citet{heesen2011}, who studied IC~10 with radio continuum polarimetry, to estimate the possible RM contribution of a starburst dwarf galaxy. 
We assume a constant thermal electron density $n_e$ in the galactic mid-plane, which falls off exponentially with height. 
For the magnetic field, a combination of a spiral plane-parallel field and a poloidal vertical field both with a characteristic strength $B_{\rm host}$ is used.
We neglect random components of the magnetic field since they do not significantly affect the distribution of $\text{RM}$.
The distribution of stars in dwarf galaxies is centered on the disc.
We model their distribution with an exponential with a scale height of 300 pc \citep[similar to][who studied IC~10]{leroy2006IC10}.

\subsubsection{Likelihood function}
Within the MW-like spiral galaxy, we draw a sample of possible positions of the progenitor, according to the assumed distribution function.
Tests showed that 1000 positions are enough to ensure converged results.
For each of these positions, we compute the full sky of $\DMHost$ and $\RMHost$ measurements, similar to the approach used by \citet{Walker2018}.
The LoS integral is then computed to the edge of the host in all different directions defined by the {\sc HEALPIX} \citep{Gorski2005} tessellation of the sphere.
The probability density of values on the full-sky delivers the likelihood functions $P(\DMHost)$ and $P(\RMHost)$.
The sum of the likelihood functions at the different positions is then the full likelihood function for the host galaxy. \\

Note that the results at the position of the Sun can be used to obtain predictions for the contribution from the MW itself.
By construction, the results are identical to results in \citet{jf12}. \\

For the starburst dwarf galaxy, we compute LoS integrals for different inclination angles and penetration depths, such that the assumed distribution of FRBs in the host is constant throughout the disc.
Since the model is rotationally invariant, variations of the azimuthal angle are redundant.
LoS are calculated until they leave the disc, excluding contributions of the galactic halo, which, however, is expected to be at least 1 order of magnitude below the galactic contribution. 

To account for possible variance across the distribution of similar starburst dwarf galaxies, we combine predictions for several choices of $n_e$ and $B_{\rm host}$, according to prior distributions explained in detail in App. \ref{sec:priors}. \\

The dispersion delay produced at the host increases, due to cosmic expansion.
The observed contribution of $\DMHost$ to the total $\DMobs$ scales with the source redshift as $(1+z)^{-1}$ \citep[e. g.][]{Macquart2015}, so the likelihood function shifts accordingly \citep{Walker2018} as
\begin{equation}
p(\DMHost|z_s) = (1+z_s) p((1+z_s) \DMHost|z_0) .
\label{eq:DM-redshift-likelihood}
\end{equation}
For $\text{RM} = \frac{\Delta \Phi}{\Delta \lambda^2}$, the contribution of the host scales with $(1+z)^{-2}$ instead.
Therefore, the corresponding likelihood function shifts as
\begin{equation}
p(\RMHost|z_s) = (1+z_s)^2 p((1+z_s)^2 \RMHost|z_0) .
\label{eq:RM-redshift-likelihood}
\end{equation}

\subsection{Local Environment}
\label{sec:progenitor}
\subsubsection{Model}
We assume FRBs to be produced by magnetars \cite[e. g.][]{Popov2010,Pen2015,Katz2016, Metzger2017,Beloborodov2017,Metzger2019}.
\new{Neutron stars are generally considered one of the most likely sources for FRBs.}
\citet{Beniamini2019} concluded that 12--100 per cent of neutron stars are born as magnetars.
Hence, it is expected that they are numerous around star-forming regions.
Their number density scales with the star formation rate.
Results of \citet{Niino2018} and \citet{locatelli2018} suggest that the number density of FRBs does also scale with the star formation rate.
This makes magnetars a likely candidate for the source of FRBs. 
\new{Many other objects have been proposed as sources of FRBs \citep[see][who provide a living catalog of theories]{platts2018livingcat}.
    We restrict this study to exemplary compare two models of the local environment of the FRB progenitor.
    }

To account for the local environment of a magnetar FRB progenitor, we make use of the models and results of \citet{Piro2018}.
They give theoretical predictions for the $\DMProg$ and $\RMProg$ from the local environment of the FRB, assuming they are produced by a young neutron star.
They consider two models.
One model assumes a uniform local ISM, while the other accounts for changes in the ISM due to stellar winds of the seed star.

In this work, we consider the two models for the \uniform and \wind cases, plus we consider the additional contribution of the SNR environment for the latter model in the \windSNR case.
We use this low number of models to show how multiple progenitor models can be compared and combined to be tested against observations. \\

Stellar winds cause the magnetic field of the local environment to align and form a significant large-scale component.
\new{The RM-predictions for that environment in the \wind model are thus much more robust than for the supernova remnants.
    The latter model assumes the shock-generated magnetic field to be coherent, while the topology is very likely random.
    Hence, results for the \uniform and the \windSNR model should be considered as upper limits.}

\subsubsection{Likelihood function}
Under the assumption that FRBs are produced at young neutron stars, \citet{Piro2018} have derived expectation values for the $\DMProg$ and $\RMProg$ of the local FRB environment.
These are given as functions of the ISM number density $\nISM$, the time since the SN $t$, the energy of the explosion $E$, the mass of SN ejecta $M$, the wind mass loading parameter $K$, the stellar radius $R_\star$ and the stellar magnetic field $B_\star$. \\

The likelihood function is obtained with a Monte-Carlo method, where we sample these parameters with reasonable prior distributions, calculate the corresponding $\DMProg$ and $\RMProg$ and compute their probability density.
The priors chosen to obtain those are summarized in App. \ref{sec:priors}. 

The contribution from the progenitor undergoes the same evolution with redshift as the contribution from the host, see Eqs. \ref{eq:DM-redshift-likelihood} \& \ref{eq:RM-redshift-likelihood}

\section{Model Results}
\label{sec:model_results}

\subsection{IGM, Constrained results for the local Universe}
\subsubsection{Dispersion Measure}
\begin{figure}
    \centering
    \includegraphics[width=0.5\textwidth]{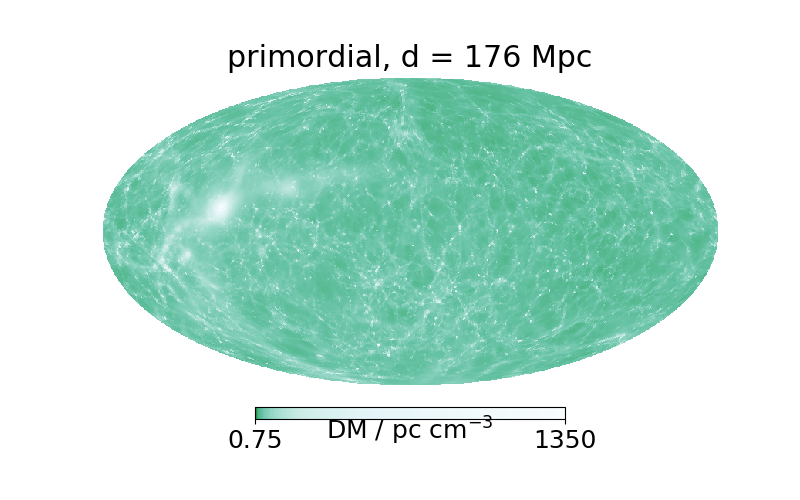}
    \caption{Full-sky map of $\DMIGM$ predictions for sources at 176 Mpc distance in the local Universe. Results are shown in super-galactic coordinates for the \primordial model.
    The distribution of free electrons, hence $\text{DM}$, is identical to the \astrophysical case.
    }
    \label{fig:constrained_DM_skymap}
\end{figure}
\begin{figure}
    \centering
    \includegraphics[width=0.5\textwidth]{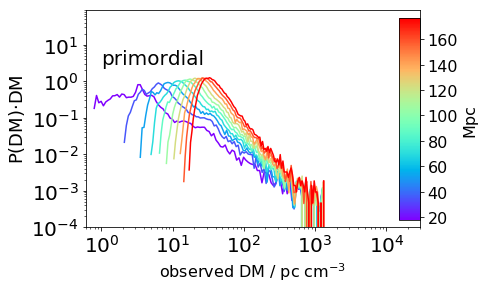}
    \caption{Likelihood function $P(\DMIGM|d)$ for FRB sources at distance $d$ in the local Universe, $d \lesssim 176$ Mpc, for the \primordial model.
    The distribution of free electrons, hence $\text{DM}$, is identical to the \astrophysical case.
    }
    \label{fig:likelihood_DM_IGM_constrained}
\end{figure}

In Fig. \ref{fig:constrained_DM_skymap} we show the full-sky projection of the expected $\DMIGM$ of FRBs at a distance of 176 Mpc.
This nicely shows the distribution of structure in the local Universe (see \citealt{Hackstein2018}). 
The Virgo cluster is the most dominant local contributor with up to  $\DMIGM \gtrsim 10^3 \unitDM$.

The $\DMIGM$ prediction is the same in both IGM models, as they result in almost identical distribution of gas. \\
Taken from such full-sky maps at different redshift, in Fig. \ref{fig:likelihood_DM_IGM_constrained} we present the evolution of the likelihood function of $\DMIGM$.
These results agree reasonably well with results in \citet{dolag2015} and \citet{Walker2018}.
At short distances, the tail at high values is more pronounced, caused by nearby, high-density regions.
With increasing distance, the distribution moves towards a log-normal distribution, the mean of which increases steadily due to the cumulative growth of $\DMIGM$.
Also, an increasing number of LoS crosses high-density structures, which add to the tail at high values.

\subsubsection{Rotation Measure}
\begin{figure}
    \centering
    \includegraphics[width=0.47\textwidth]{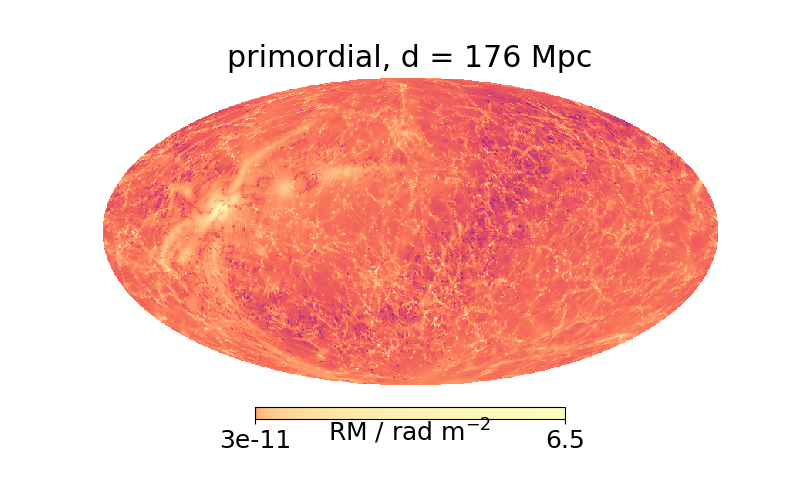}
    \includegraphics[width=0.47\textwidth]{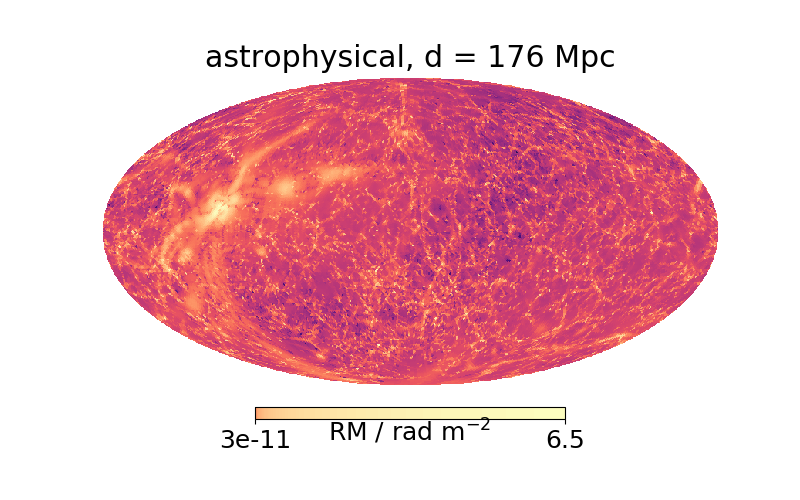}
    \caption{Full-sky map of $|\RMIGM|$ predictions for sources at 176 Mpc distance in the local Universe, for the \primordial (top) and \astrophysical model (bottom). 
    Results are shown in super-galactic coordinates. }
    \label{fig:constrained_RM_skymap}
\end{figure}
\begin{figure}
    \centering
    \includegraphics[width=0.47\textwidth]{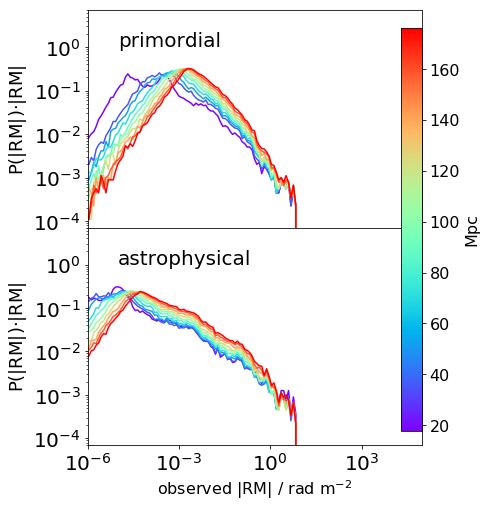}
    \caption{Likelihood function $P(|\RMIGM||d)$ for FRB sources at distance $d$ in the local Universe, $d \lesssim 176$ Mpc for the \primordial (top) and \astrophysical model (bottom).
    }
    \label{fig:likelihood_|RM|_IGM_constrained}
\end{figure}
In Fig.~\ref{fig:constrained_RM_skymap} we show the full-sky projection of expected $\RMIGM$ of FRBs at a distance of 176 Mpc for, both, the \primordial and the \astrophysical model.
The structure of the local Universe is nicely reproduced.
Again, the Virgo cluster appears as the most dominant contributor with up to $\RMIGM \gtrsim 6 \unitRM$, which roughly agrees with the observations of \citet{vallee1990}.

Both IGM models result in almost identical maximum values of $\RMIGM$.
These are found in LoS that pass through regions of high density, like the Virgo cluster, that contribute very high values of $\RMIGM$.
The models were built to reproduce the conditions observed in these regions in the local Universe.
However, the two magneto-genesis scenarios result in severely different magnetic fields in voids.
Fig. \ref{fig:constrained_RM_skymap} shows, LoS that do not pass through regions of high density have $\RMIGM$ lower by up to two orders of magnitude. \\

Taken from such full-sky maps at different redshifts, in Fig. \ref{fig:likelihood_|RM|_IGM_constrained} we present the evolution of the likelihood function of $\RMIGM$.
Since the distribution of positive and negative values is very similar, we make use of $\log(|\text{RM}|)$ in all our likelihood functions to compare contributions of different order in more detail. 

The highest values, $ \RMIGM \approx 1-10 \unitRM$, agree in both models.
These are LoS that intersect high-density regions, associated with the $\rho/\langle \rho \rangle \geq 10^2$ overdensity of galaxy clusters,  contributing high values of $\RMIGM$. 
However, the fraction of such LoS is limited, and they do not affect 
much the overall distribution of $\RMIGM$ \citep{Vazza2018_FRB}.

As the peak increases with distance, the \astrophysical case peaks about two orders of magnitude below the \primordial case.
However, the overall contribution of $\RMIGM$ is much too low to have significant influence on the total $\RMobs$ within maximum distance in the constrained volume, 176 Mpc.
This also holds for possibly different results for positive and negative $\RMIGM$ caused by dense structure outside of cores of clusters. \\

Note that the \primordial model stasrted from a magnetic field that was coherent over the whole simulation volume.
Outside of dense structures, this topology of the initial field is conserved and results in very optimistic estimates of $\RMIGM$, as contributions from separate parts of the LoS cannot cancel each other.
A more detailed study of this effect can be found in App. \ref{app:uniform}.
Note that for the constrained distance, this feature is of order $10^{-2} \unitRM$ in the \primordial case, subdominant to other contributions along the LoS and hence not observable.
At greater distances, we combine separate trajectories with random orientations, thus enabling the contributions from separate sections of the LOS to cancel each other.

\subsection{IGM, High redshift results}
\subsubsection{Dispersion Measure}
\begin{figure}
    \centering
    \includegraphics[width=0.47\textwidth]{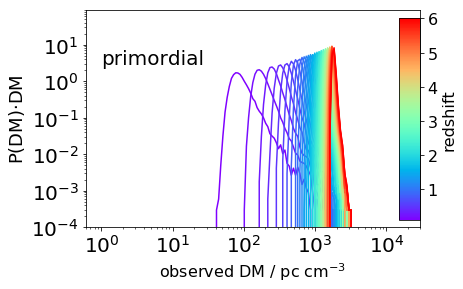}
    \caption{Likelihood function $P(\DMIGM|z)$ for FRB sources at redshift $z$ in the distant Universe for the \primordial model.
    The distribution of free electrons, hence $\text{DM}$, is identical to the \astrophysical case.
    From blue to red, the graphs show results at redshifts $z=0.1$ to $z=6.0$ in steps of 0.1.
    }
    \label{fig:likelihood_DM_IGM}
\end{figure}
In Fig. \ref{fig:likelihood_DM_IGM} we present the resulting likelihood function of $\DMIGM$ contribution from the IGM for FRB at different redshifts in the distant universe for the \primordial model.
The distribution of free electrons, hence $\text{DM}$, is identical to the \astrophysical case.
The distribution of $\DMIGM(z=1)$ is very peaked around $1000 \unitDM$, in good agreement with results of previous studies, where this value is reported to be $855 - 1200 \unitDM$ \citep{ioka2003,inoue2004,McQuinn2013, Deng2014,dolag2015,Walker2018,Pol2019}.
The shape is similar to results in Fig. \ref{fig:likelihood_DM_IGM_constrained} at the highest distance and compares well with the results of \citet{dolag2015} and \citet{Walker2018}.

With increasing redshift, the proper density of free electrons increases,
as does the average $\DMIGM$ contribution of the IGM.
This makes the whole likelihood function $P(\DMIGM)$ shift to higher values with increasing redshift.
As the cumulative growth of $\DMIGM$ from low-density regions approaches the scale of dense structure contributions, $P(\DMIGM)$ becomes much narrower .
However, the overall change is slower at higher redshift $z$ \citep{zheng2014}. 
Therefore, the likelihood function for high $\DMIGM$ is much broader in $z$.
This shows that, although the $\text{DM}$ delivers good upper limits on $z$, the uncertainties in the estimate will always remain rather large and other ways to infer $z$, e. g. by identification of the host, are preferred \citep[cf.][]{dolag2015,Walker2018,Kumar2019,Pol2019}.
\subsubsection{Rotation Measure}
\begin{figure}
    \centering
    \includegraphics[width=0.47\textwidth]{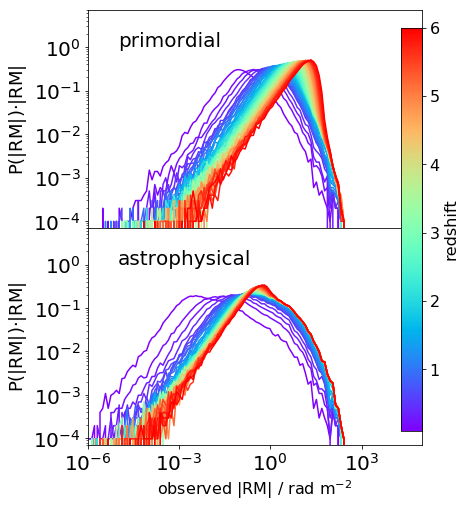}
    \caption{Likelihood function $P(\RMIGM|z)$ for FRB sources at redshift $z$ in the distant Universe in a \primordial (top) and \astrophysical (down) model.
    From blue to red, the graphs show results at redshifts $z=0.1$ to $z=6.0$ in steps of 0.1. 
    }
    \label{fig:likelihood_RM_IGM}
\end{figure}
In Fig.~\ref{fig:likelihood_RM_IGM} we present the likelihood function of $\RMIGM$ contribution from the IGM for FRB at redshift $z$ in the \primordial and \astrophysical models of the distant Universe.
As the models used here were produced with the same tools and physics as the ones used by \citet{Vazza2018_FRB}, the results we find are quite similar.
However, the average value of these distributions is significantly lower than the results of \citet{Akahori2016}, which is due to the lower magnetic field strength outside of clusters.
Here we use $B \sim 0.1$ nG instead of the $10 - 100$ nG of \citet{Akahori2016}, due to the more efficient dynamo amplification assumed in the latter model.  \\

At lower redshifts, $z\sim 0.1$, $\RMIGM$ tends to low values close to zero.
Only a few LoS show high values of up to $\RMIGM \sim 100$ rad m$^{-2}$.
These are LoS that traverse high-density regions, associated with the $\rho/\langle \rho \rangle \geq 10^2$ over-density of galaxy clusters, which contain amplified magnetic fields.
With higher redshift, more and more LoS traverse clusters, some even twice, and their $\RMIGM$ reach values above 100 rad m$^{-2}$ in both the \primordial and the \astrophysical cases.

Many of the LoS traverse the low-density IGM only and contribute most of $\RMIGM$.
The cumulative growth shifts $\RMIGM$ to higher values, but slower than $\DMIGM$, as $\RMIGM$ from different regions of the IGM can cancel each other.

The IGM model we used considers an initial magnetic field that is coherent over 250 Mpc/h, i. e. the full simulation volume.
This is well conserved in low-density regions and results in a uniform sign of $\RMIGM$ contributions along a continuous LoS segment.
However, since several of these segments with random orientation are combined to obtain the full LoS, they can cancel each other and we obtain results that are statistically equivalent to a stochastic field with lower coherence length. \\

There is a significant difference in $P(\RMIGM|z)$ between the \primordial and \astrophysical cases.
The peak of $\RMIGM$ is 2 orders of magnitude lower in the latter case, similar to results at low redshift, shown in Fig. \ref{fig:likelihood_|RM|_IGM_constrained},
Further, the shape looks increasingly different at higher redshift $z$.
Though the peak value is rather low, $\lesssim 10 \unitRM$ still at $z=6$, the different shapes will likely reflect in the distribution of extragalactic $\RMEG$, given that there is no dominant contribution from the other regions.

\subsection{Progenitor environment, host galaxy and MW}
\begin{figure*}
    \centering
    \includegraphics[width=0.7\textwidth]{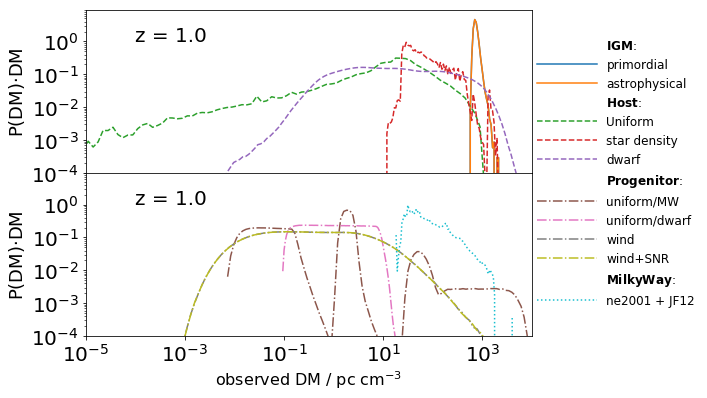}
    \caption{Likelihood functions $P(\text{DM}|z=1)$ for all contributor models investigated in this work.
    The linestyle indicates the contributing region described by the model. 
    }
    \label{fig:likelihoods_DM}
\end{figure*}
\begin{figure*}
    \centering
    \includegraphics[width=0.7\textwidth]{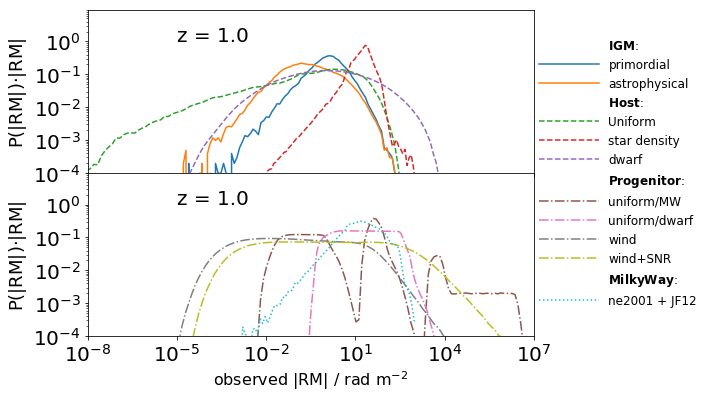}
    \caption{Likelihood functions $P(\text{RM}|z=1)$ for all contributor models investigated in this work.
    The linestyle indicates the contributing region described by the model.
    }
    \label{fig:likelihoods_RM}
\end{figure*}
\subsubsection{Dispersion Measure}
The likelihood functions of $\text{DM}$ for all models investigated in this work are presented at redshift $z=1$ in Fig. \ref{fig:likelihoods_DM}.

The two models for the IGM, \primordial and \astrophysical, have identical $\DMIGM$ by construction.
The two behaviours overlap each other.
The dominant peak is at around $10^3 \unitDM$. \\

The model that assumes a spiral host galaxy similar to the MW is modelled with two distribution functions of the position of the FRB progenitor, one is \Uniform, the other follows the \stardensity in the MW.
The bulk of both of these distributions is similar to the MW's.
There is less $\DMHost$ around $10^3 \unitDM$, since there are less LoS that traverse big parts of the galaxy.
For the \Uniform distribution, a lot of progenitors are located close to the border of their host.
A huge number of LoS traverse only small parts of the galaxy.
Therefore, the tail towards lower values is much more pronounced.

\citet{Xu2015} also investigate a spiral galaxy.
The maximum, $\approx 1500 \unitDM$, and most probable value, $\approx \rm few \unitDM$, are similar to our results. \\

The likelihood function for starburst \heesen galaxies as FRB hosts, shows a flat plateau at $\DMHost = 1-10^3 \unitDM$ due to the assumed flat prior.
In most cases the contribution will be significantly lower than the IGM.
However, there is a small probability of a few per cent that it contributes more to the $\DMEG$. \\

The \uniform model described by \citet{Piro2018} strongly depends on the ISM density $\nISM$.
The shape of the likelihood function is almost identical to the chosen prior distribution $\pi( \nISM )$.
Of course, this depends on the host galaxy and we will show below the result for both host galaxy models investigated in this work.

For the case of MW-like spiral galaxies, we see in Fig.~\ref{fig:likelihoods_DM} that the supernova remnants can provide an observed $\DMProg$ up to several $10^3 \unitDM$, even at a distance of $z=1$, if the magnetar is located in an H\,{\sc ii}-region.
Only the \heesen host model has very small chance to contribute similarly high values of $\DMHost$.
None of the other models is able to produce such high values of $\text{DM}$.
This shows how likelihood functions can be used to rule out contributor models for single events and, subsequently, for whole populations. \\

Fig. \ref{fig:likelihoods_DM} also shows that a high $\DMobs$ does not necessarily imply a high redshift, but can also be produced in the local environment of the FRB, even if the probability is rather low, $\lesssim 1$ per cent.
However, if future observations reveal a significantly higher number of large $\DMobs \gtrsim 10^3 \unitDM$, this would argue in favour of a population at reasonable cosmic distance, $z \gtrsim 1$. \\

The \wind model in \cite{Piro2018} results in a rather flat distribution of $\DMProg$ around $10^{-2} - 10^{-1} \unitDM$ that decreases rapidly at both ends.
Adding the SNR contribution in \windSNR, the plateau expands to substantially higher values of $10^{1} \unitDM$ and the tail includes $\DMProg \gtrsim 10^2 \unitDM$ with probability of $\approx 0.1$ per cent.
At redshift $z=1$ this is far below the IGM contribution.

\subsubsection{Rotation Measure}
Fig.~\ref{fig:likelihoods_RM} shows the likelihood functions of $\text{RM}$ for all models at redshift $z=1$.\\

The model for the MW is in agreement with the data provided by \citet{oppermann2015}. 
The likelihood function is of similar shape as for the IGM, about an order of magnitude above the \primordial model.
It stays above both models of the IGM for all redshifts probed in this work, $z \leq 6$. \\

The host model that resembles a MW-like spiral galaxy shows a likelihood function for $\RMHost$ that is very peaked around $10^1 - 10^2 \unitRM$ -- about a magnitude above the peak of the \primordial model -- in case the positions of FRB progenitors scale with the {\it star density}.
This falls off exponentially with distance from the center of the galaxy, which hosts most candidate locations and gives the strongest contribution to $\RMHost$.

For a \Uniform distribution of progenitors, there is a wide and pronounced tail towards lower values of $\RMHost$, due to the numerous short LoS of progenitors located at the border of the galaxy.
In this case, the bulk of $\RMHost$ is comparable to the IGM contribution. 
These models are in best agreement with the results by \citet{Basu2018}, who investigated the $\text{RM}$ contribution of a randomly orientated galaxy in the LoS of a quasar. 
The range up to $\lesssim \rm few\ 100 \unitRM$ and median $\approx 10 \unitRM$ of their distribution is comparable to our results.

The starburst \heesen galaxy model assumes the distribution of progenitor positions to be concentrated close to the galactic centre.
Overall, the contribution is stronger than for a \Uniform distribution of sources in a spiral galaxy, since most LoS traverse considerable portions of high density regions in the galactic disc. Due to the small size of a dwarf galaxy, the majority of LoS show $\RMHost$  below the most probable value found for \stardensity distribution in spiral galaxies. \\

The \uniform model of the local environment of neutron stars described in \cite{Piro2018} strongly depends on the local ISM density $\nISM$.
Hence, the shape of the likelihood function is determined by the prior distribution chosen for $\nISM$ and allows us to easily associate $\RMProg$ with the medium around the progenitor.
This depends on the galaxy that hosts the FRB and we present results for both models of the host galaxy investigated in this work.
In case of a spiral galaxy like the MW we see that for magnetars located in H\,{\sc ii}-regions, the contribution of the remnants of the recent supernova can reach extremely high $\RMProg$ up to several $10^6 \unitRM$, exceeding $\text{RM}$s in that region observed with background sources by several orders of magnitude \citep[e. g.][]{harvey2011}.

There is a reasonable probability of about $1$ per cent to see $\RMProg \gtrsim 10^4 - 10^5 \unitRM$ from magnetars in these regions.
This suggests that the high $\RMobs$  of FRB121102 \citep{Michilli2018} might be the signal of an FRB located in an H\,{\sc ii}-region.
However, the bulk of $\RMProg$ expected in both models is of the order of the contribution of the IGM or the MW. \\

If the local environment of the magnetar was instead dominated by the stellar \wind of the seed star, the likelihood function of $\RMProg$ is rather flat $10^{-4} - 10^1 \unitRM$ with rapidly falling tails on both sides.
Adding the SNR contribution in \windSNR, the plateau expands to $10^3 \unitRM$ with a high end tail reaching out to $10^5 \unitRM$.
However, it barely reaches values high enough to explain the high $\RMobs$ of FRB121102.
Since this model is more of an upper limit than a prediction, this scenario is highly unlikely.
Therefore, the best fit scenario for FRB121102 from the models of this paper is a magnetar localized in an H\,{\sc ii}-region.
This is in close agreement with previous works, which concluded that FRB121102 is likely produced by a magnetar in high density regions  \citep{masui2015,spitler2016,Beloborodov2017}.
Note, however, that a wide range of sizes and densities of H\,{\sc ii}-regions are excluded by constraints from $\text{DM}$ and the abscence of free--free absorption \citep{Michilli2018}.

\subsection{Dependence on redshift }
\label{sec:redshift_dependence}
\begin{figure*}
    \centering
    \includegraphics[width=0.6\textwidth]{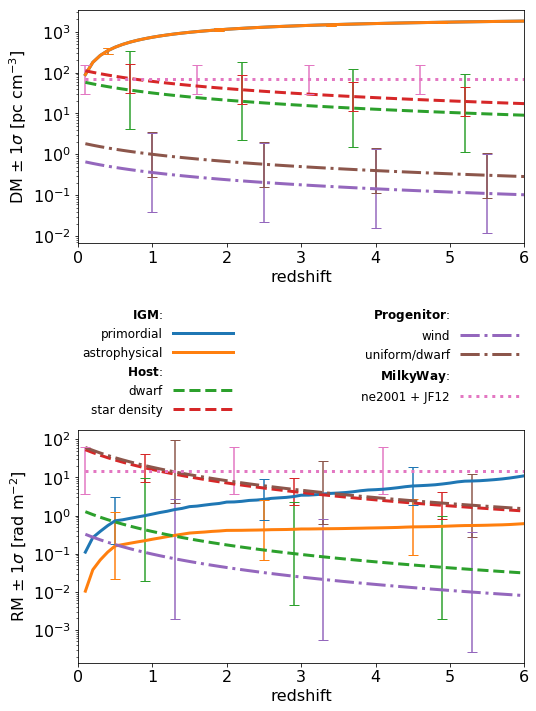}
    \caption{Redshift dependence in the distant Universe of the different average contributions $\avg{\text{DM}}$ (top) and $\avg{\text{RM}}$ (bottom).}
    \label{fig:redshift_dependence}
\end{figure*}
From the likelihood functions derived above, we compute the expectation value and deviation of the contributor models in order to compare their contribution at different redshift more easily.
The results are shown in Fig.~\ref{fig:redshift_dependence}. \\

The MHD simulations probed by \citet{Vazza2018_FRB} are produced in the same framework.
We use similar starting parameters, adding the constrained initial conditions.
The resulting LoS are, statistically speaking, almost identical.
The redshift dependence of the average $\avg{\text{DM}}$-contribution  of the IGM compares well to the results of \citet{Akahori2016}.
Since the other extragalactic contributions decrease with redshift, the IGM strongly dominates the total $\DMobs$ at redshifts $z \gtrsim 1$.
However, there is little change in $\DMIGM$ with redshift $z > 1-2$.
This introduces huge uncertainties in estimating the corresponding redshift for high $\text{DM}$s.  \\

At low redshifts, $z<0.1$, the IGM contribution is substantially sub-dominant to the contributions of the MW and the host galaxy.
Hence, the $\DMobs$ can only provide an upper limit on $z$ \citep[cf.][]{dolag2015,Niino2018,Walker2018,Pol2019}.
The different models for progenitor environment and host galaxy do not show significant differences. \\

For the $\avg{\text{RM}}$, the different models of progenitor environment and host galaxy result in rather different contributions.
E. g. a spiral galaxy similar to the MW on average contributes two orders of magnitude higher $\RMHost$ than a dwarf galaxy similar to IC~10. \\

Regardless of the model, the contribution from the host galaxy and/or the progenitor environment dominates the $\RMobs$ of FRBs in the local Universe $z < 0.1$.
The choice of models determines at which point the IGM will take over.
Although the contribution of the MW is dominant at all redshifts up to $z=6$, we argue that this contribution can be removed by subtracting the MW component with sophisticated modelling of the Galaxy \citep{imagine_whitepaper}.
At high latitudes, $\RMMW \approx 10 \unitRM$ are still very likely. 
Hence, it does not suffice to restrict the sample to FRBs observed outside the Galactic plane. \\

The difference in average $\avg{\RMIGM}$ between the primordial and astrophysical models is about one order of magnitude at $z=1$.
That difference increases with redshift to almost two orders of magnitude at $z=6$, where the \primordial model is dominant over all other extragalactic contributions.
This shows that $\RMEG$ of FRBs deliver information on and can be used to constrain the strength and origin of the IGMF.
However, the minimum redshift of FRBs required to allow us to derive conclusions strongly differs for different host galaxies and progenitor environments.

\section{Combined Results}
\label{sec:combine}

\subsection{Extragalactic likelihood function}
\begin{figure}
    \centering
    \includegraphics[width=0.5\textwidth]{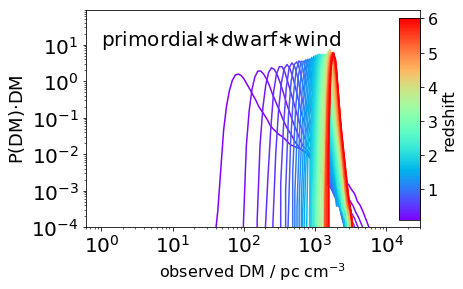}
    \caption{Combined likelihood function $P_{\rm EG}$ of all extragalactic contributors to $\text{DM}$, assuming that FRBs are produced at redshift $z$ in a \wind progenitor environment
    and hosted by a starburst \heesen galaxy
    embedded in an IGMF of \primordial origin.
    From blue to red the graphs show results for increasing redshift in the distant Universe, $0.1 \leq z \leq 6.0$ in steps of 0.1.
    }
    \label{fig:full_likelihood_DM}
\end{figure}
\begin{figure}
    \centering
    \includegraphics[width=0.5\textwidth]{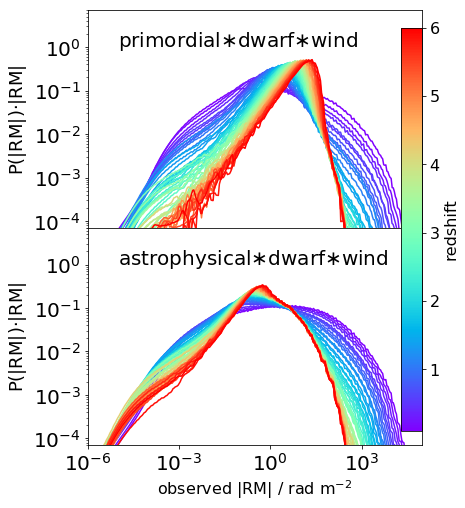}
    \caption{Combined likelihood $P_{\rm EG}$ of all extragalactic contributors to $\text{RM}$, assuming that FRBs are produced at redshift $z$ in a s\wind progenitor environment
    and hosted by a starburst \heesen galaxy 
    embedded in an intragalactic magnetic field of \primordial (top) or  \astrophysical (bottom) origin.
    From blue to red the graphs show results for increasing redshift in the distant Universe, $0.1 \leq z \leq 6.0$ in steps of 0.1. }
    \label{fig:full_likelihood_RM}
\end{figure}
\begin{figure}
    \centering
    \includegraphics[width=0.45\textwidth]{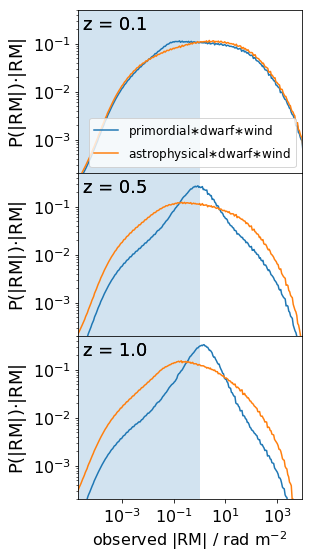}
    \caption{Same as Fig. \ref{fig:full_likelihood_RM} with both models in a single plot for a small set of redshifts to allow better comparison.
    The gray area indicates $\RMEG < 1 \unitRM$ that we consider as not observable due to uncertainties in removing the foreground of the MW and ionosphere.
    }
    \label{fig:full_likelihood_compare}
\end{figure}

In the previous sections we derived likelihood functions for all extragalactic contributors of $\DMobs$ and $\RMobs$ measured for FRBs.
In this section we combine these results into a likelihood function for the total extragalactic contribution.
The combined likelihood function of the sum of independent variables is the convolution of their individual likelihood functions,
\begin{equation}
    P_{\rm EG} =  P_{\rm IGM} \ast P_{\rm host} \ast P_{\rm prog} .
\label{eq:full_likelihood}
\end{equation}
\new{We stress that the results presented in this section can not yet be compared to observations directly, without assumptions on the FRB population and observational selection effects.
    If for example the number of FRBs increases with redshift, higher values of DM and RM are expected than for a constant number of FRBs.
    In the future, population assumptions and selection effects will be implemented using results of {\sc FRBpoppy}\footnote{\href{https://github.com/davidgardenier/frbpoppy}{https://github.com/davidgardenier/frbpoppy}} in order to provide detailed predictions, tailored to the individual telescope, to be compared to observations.
} \\

We compute the likelihood of the extragalactic component $\DMEG$, assuming that FRBs are produced in a \wind progenitor environment hosted by a starburst \heesen galaxy.
This set of models was chosen in order to obtain the most optimistic results on obtaining info about the IGMF.
Since the density distribution is the same in the \primordial and \astrophysical IGM models, we show results for the former, only.
These are shown in Fig. \ref{fig:full_likelihood_DM} for FRBs at different redshifts.\\

As explained in the previous section, the $\DMEG$ is strongly dominated by the IGM at high redshifts $z>1$.
Therefore, the combined likelihood function is almost identical to that of the IGM alone.
The distribution in Fig.~\ref{fig:full_likelihood_DM} becomes much narrower.
The range reduces from over two orders of magnitude, $\sim 10^2 - 10^4 \unitDM$, at redshift $z=0.1$ to a range of less than factor 2 at redshift $z=6$, peaked at around $2\cdot 10^3 \unitDM$.
The peak value is determined by the IGM and increases with redshift.
The tail to high $\RMEG$, provieded by strong progenitor contribution, decreases and is completely overshadowed by the IGM distribution by redhsift $z\approx 1$.
However, the increase of the peak value is rather slow at high redshift.
This introduces a high uncertainty in determination of the exact redshift using $\DMobs$. \\

The contribution of the host galaxy can cause huge values of observed $\DMHost$. which exceed the contribution of the IGM even at very high redshifts, $z\gtrsim 6$.
Therefore, high $\text{DM}$s do not necessarily imply a high redshift of the source, but could also be produced in a nearby host galaxy.
Note, though, that the likelihood of high $\DMHost$ at low $z < 1$ is rather low, $\lesssim$ few per cent, as the bulk of $\DMHost$ is about an order of magnitude below results of the IGM at $z>1$.
If the observed amount of FRBs with high $\text{DM}$s is found to be $\gtrsim 5$ per cent, this would allow to conclude on a cosmic population $z>1$. \\

We further compute the likelihood of the extragalactic component $\RMEG$, assuming that FRBs are produced in a {\it wind} progenitor environment hosted by a starburst dwarf galaxy.
To see the difference for the scenarios of magneto-genesis of IGMFs, we compute results for, both, the \primordial and \astrophysical cases.
The results for FRBs at different redshift are shown in Figs. \ref{fig:full_likelihood_RM} \& \ref{fig:full_likelihood_compare}.

At low redshift, the shape of $P(\RMEG|z)$ is determined by the host contribution.
However, there is a significant difference between the two models, already at $z=0.5$, that grows with redshift, though the average of both distributions is comparable.
A quantification of that difference can be found in Sec. \ref{sec:fake_test}.

For the \primordial model, contributions from the IGM become comparable to the host contribution at $z \approx 0.5$.
This allows to lower the chance of the highest $\RMEG$ due to cancellation of $\text{RM}$ from different regions, while intermediate results $\gtrsim 1 \unitRM$ become more likely.

At higher redshift, $z\gtrsim 4$, the shape is completely determined by the IGM contribution, as it exceeds the observed contribution of the host galaxy at such high redshifts.
This shows the capability of $\RMobs$ of FRBs to shed light on the origin of IGMFs. \\

Note that, although the values of $\RMIGM$ in the \astrophysical case are equal or smaller than in the \primordial case, there can be a slightly higher chance of a high $\RMEG$ in the former case.
This is because $\text{RM}$s from different regions of the LoS, e. g. IGM and host, can cancel each other.
Hence, two comparably strong contributors can weaken the likelihood for high $\RMobs$, as compared to only one dominant contributor.
Use of the likelihood function can account for that, which is an advantage as compared to considering only the average value. \\

We stress that results in this section highly depend on the choice of contributor models.
Here, we made use of those host galaxy and progenitor environment models, which showed the least contribution to $\RMEG$.
We did this in order to derive the most optimistic results on obtaining info on the IGM.
The results in Figs. \ref{fig:likelihoods_DM}, \ref{fig:likelihoods_RM} and \ref{fig:redshift_dependence} show that the other host and progenitor models investigated here provide much higher values of $\text{RM}$ that overshadow the IGM contribution up to redshift $z=3-4$.
Ways to restrict the inference to those FRBs that fit the presented choice of models will be discussed in Sec. \ref{sec:discussion} and will be the subject of upcoming works.

\subsection{Application to observations}
\label{sec:fake_test}

\begin{figure}
    \centering
    \includegraphics[width=0.47\textwidth]{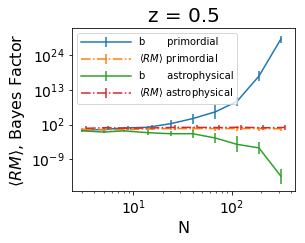}
    \caption{Bayes factor $b$ and average $\avg{\RMEG}$ for two fake populations at the indicated redshift that resemble the \primordial and \astrophysical case.
    The errorbars of $\avg{\RMEG}$ show the $1\sigma$ standard deviation of $\text{RM}$ in the population.
    For $b$ they show the standard deviation of 6 random samples of the population.}
    \label{fig:fake_test}
\end{figure}
\begin{figure*}
    \centering
    \includegraphics[width=\textwidth]{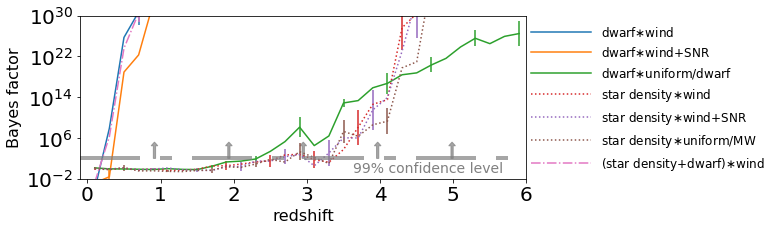}
    \caption{Bayes factor for fake samples of 100 FRBs at different redshift, resembling the population in the \primordial magneto-genesis scenario, combined with several sets of models indicated by colors and linestyles.
    The solid lines consider a \heesen galaxy as the host of FRBs, 
    the dotted lines assume the MW-like spiral galaxy, \stardensity, 
    while the dash-dotted line allows for both of these galaxy types to host similar numbers of FRBs.
    The gray line marks a Bayes factor of $10^2$, that indicates a 100-times higher chance for the fake population to be produced in the \primordial, rather than in the \astrophysical scenario, assuming the same models for the other contributors.
    This indicates the 99 per cent confidence level to rule out the latter scenario in favor of the former.
    }
    \label{fig:faketests_redshift}
\end{figure*}

At this point, there are few observations of FRBs with $\RMobs$. 
This will change soon, after new telescopes dedicated to observe FRBs, e. g. CHIME/FRB, FAST, SKA and MeerKat, begin producing $\text{RM}$ data \citep{Jonas2009,Nan2011,Keane2013,Macquart2015,CHIME/FRB2018}.
We therefore investigate samples of random tuples of $\text{DM}$ and $\text{RM}$ that resemble the expected distribution at redshift $z=0.5$ shown in Figs. \ref{fig:likelihood_DM_IGM} and \ref{fig:full_likelihood_compare}.
Note that contributions from the ionosphere to the $\text{RM}$ is expected to be a few$\unitRM$ \citep{Weisberg2003}, therefor hampers investigation of the distribution of $\RMobs < 1\unitRM$.
To account for that, we only sample $\text{RM}$ above that value. 

For each of the ($\text{DM}$,$\text{RM}$)-tuples, we compute the Bayes factor 
\begin{equation}
    b(\text{DM},\text{RM}|{\rm primordial,astrophysical}) 
    =
    \frac{ P(\text{DM},\text{RM}|{\rm primordial}) }{ P(\text{DM},\text{RM}|{ \rm astrophysical}) },
\label{eq:bayes-factor}
\end{equation}
which quantifies how much more likely it is that the given tuple of $\DMEG$ and $\RMEG$ is produced in the \primordial rather than in the \astrophysical case.
A Bayes factor $b(O, M_1,M_2) > 10$ shows that observation $O$ is more than 10 times more likely in model $M_1$ than in model $M_2$.
This signals strong evidence in favor of $M_1$ as compared to $M_2$.
For values $b>100$ the evidence is considered to be decisive \citep{jeffreys1961}.

The likelihood of two events is the product of their individual likelihoods.
The same holds for the Bayes factor, that applies a number to our corroboration towards one model over the other.

In particular, we use the $\DMEG$ to derive a likelihood function $P(z|\DMEG)$ for the redshift of the FRB.
This is then used as a prior to derive the likelihood of the $\RMEG$
\begin{equation}
    P(\DMEG,\RMEG|BO) \propto \int P(\RMEG|BO,z) P(z|\DMEG) \dd z
\end{equation}

Note that this inference does not require knowledge on the redshift of the FRB, but only uses the $\DMEG$ and $\RMEG$.
If the redshift is known, $P(z|\DMEG)$ can be replaced by a narrower function in order to decrease the range of possible $\RMEG$ in the different models and allow for more precise results.
 
Note that $\text{RM}$ from active galactic nuclei are much easier to be associated with a redshift.
Hence, including AGN in our analysis in future work will significantly improve the results of this section, despite the missing $\text{DM}$. \\

We compute the Bayes factor for different sizes of the sample to see how many FRBs are required at a given redshift in order to allow conclusions on the IGM.
We also compute the average $\avg{\RMEG}$ of these samples to compare the efficiency of the Bayesian inference to the frequentists' approach.
The results are shown in Fig.~\ref{fig:fake_test}.

The $\avg{\RMEG}$ agree within $1\sigma$ standard deviation for both populations.
Whereas the Bayes factor shows a difference $\gtrsim$ 10 orders of magnitude for a number of 100 FRBs at redshift $z=0.5$, in case they are produced by magnetars in wind environments hosted by dwarf galaxies.
This huge difference clearly sets apart the scenarios for the generation of the IGMF. \\

We stress that this result is largely dependent on the assumed model for the host galaxy and progenitor environment since Fig.~\ref{fig:redshift_dependence} shows that other choices lead to very different results.
This can shift the required redshift of FRBs, e. g. hosted by spiral galaxies, to much higher redshifts, $z \gtrsim 3$ or even above.
Hence, an identification of the host galaxy as well as the local environment of the progenitor is crucial for their use to probe IGMFs. This can be a difficult task, especially for the case of dwarf galaxies \citep[e. g.][]{eftekhari2017}.\\

We compare several possible combinations of models to investigate the redshift of FRB sources required to obtain information on the IGMF in those scenarios.
We sample $\DMEG$ and $\RMEG$ of 100 FRBs, all at the same redshift and compute the corresponding Bayes factor $b(\text{DM},\text{RM}|\rm primordial,astrophysical)$ (Eq. \ref{eq:bayes-factor}).
This procedure is then repeated with increasing redshift. 
We compute six random samples at each of these redshifts and plot the average and standard deviation of the Bayes factor.
The results are shown in Fig. \ref{fig:faketests_redshift}. \\

For three of the seven combinations of models, at redshifts $z \gtrsim 0.5$, the resulting Bayes factor is $b \gg 10^2$ and hence clearly speaks for a \primordial origin of IGMFs (in case of a \primordial fake population. The same holds for the \astrophysical scenario.)
These are the combinations that assume the \wind or \windSNR model for the progenitor together with a \heesen host galaxy. 
The \wind model delivers a distribution of $\RMProg$ that is more concentrated on lower values as compared to the \uniform model.
Though the former delivers much higher values of $\RMProg$, this is mostly for times $t\lesssim 25 \rm\ yr$, below which the radio bursts are weakened by the supernova ejecta \citep{margalit2018}.
At later times, the predicted $\RMProg$ decreases much faster in the \wind than in the \uniform case, accounting for the higher amount of low $\RMProg$ in the former case. \\

The third combination considers both galaxy models, the \heesen as well as the MW-like spiral galaxy, {\it star density}, as equally likely hosts.
This is done by using the renormalised sum of both likelihood functions, shown in Figs. \ref{fig:likelihoods_DM} and \ref{fig:likelihoods_RM}.
The resulting distribution is much less peaked than the \stardensity case and tends to lower values, therefore enabling those FRBs to deliver information on the IGM.
This means that, even if not all of the FRBs taken into account are hosted by \heesen galaxies, their overall statistics at redshift $z\gtrsim 0.5$ may still allow us to draw conclusions on the magneto-genesis of IGMFs. 

We note that the equal weighting of the two host models is an arbitrary choice, not based on any realistic population synthesis of galaxies.
In reality, the weighting for different types of galaxies changes with redshift, as does the galactic stellar mass function \citep[e. g.][]{lilly2009} as well as the major star population and their age in different types of galaxies \citep[e. g.][]{hopkins2004}.
An increase of the weight of spiral galaxies, \stardensity, would increase probability of strong host contributions and hence push the redshift required to probe the IGM to higher values. Future work should account for that by assuming several possible populations of FRBs, their possible host galaxies and redshift distribution. \\

Four of the seven combinations result in undecisive Bayes factors, $b \lesssim 10^2$, at redshifts below $z\lesssim 3$.
These are scenarios that assume, either, a MW-like galaxy, {\it star density}, as the host of FRBs with any model for the progenitor investigated in this work, or, a magnetar embedded in a \uniform environment hosted by a \heesen galaxy.
In these cases, the local contribution is too strong to allow us to infer information on the IGM.
For the spiral galaxy model, the distribution of $\RMHost$ peaks at about $\gtrsim 10^1 \unitRM$ in the host rest-frame.
This strong contribution can overshadow the contribution from the IGM at high redshift.
However, even for these unfavourable models, beyond redshifts of $z \gtrsim 3.5 - 4$, the contribution of the IGM becomes strong enough to allow to distinguish between different scenarios of magneto-genesis of IGMFs.
FRBs at such high redshift probably will not even require us to select a special subset of the population, e. g. hosted by dwarf galaxies, in order to obtain reasonable results.
However, using only FRBs beyond redshift $z\sim 4$ might be even harder to accomplish, as it is a tough task, in case they exist, to find the exact redshift of FRBs at this distance.
On the one hand, the $\text{DM}$ can only be used to derive an upper limit on $z$, as we show in Sec.~\ref{sec:combine}.
On the other hand, identification of the a dwarf host galaxy becomes increasingly harder with growing redshift
\citep[e. g.][]{eftekhari2017}. \\

Note that the {\it dwarf}$\ast$\uniform case shows vastly lower values of the Bayes factor at redshifts $z>4$ than all of the \stardensity combinations.
The latter are dominated by the host contribution for all models of the progenitor environment.
The very narrow distribution $P(\RMHost|star\ density)$ peaks between the \primordial and \astrophysical $P(\RMIGM)$.
Their convolution, $P_{\rm Host}\ast P_{\rm IGM}$ is rather different for the two cases.

In contrast, the {\it dwarf}$\ast$\uniform case is dominated by the local environment of the progenitor, which shows a very flat distribution, $P(\RMProg)$, due to the assumed flat prior.
The \primordial $P(\RMIGM)$ peaks within the range of $P(\RMProg)$, their convolution, $P(\RMEG)$ has similar shape to $P(\RMProg)$, altered only by a subtle peak at high values.
The primarily low contributions of the \astrophysical $\RMIGM$ do not alter the shape of $P(\RMProg)$ significantly.
Hence, the full likelihood functions $P(\RMEG)$ have similar shape for both models of the IGM.
Single events have less weight as evidence because the Bayes factor is generally closer to unity.
In mathematical terms, the integral over the absolute difference of the likelihood functions of the two cases is higher for the \stardensity galaxy combinations than for the {\it dwarf}$\ast$\uniform case, which is hence less informative.
Note that this is another measure that might be used to infer the likelihood of different combinations of models. \\

Note further that for all four of these models, the Bayes factor $b$ drops significantly at around $z \approx 3$.
This is because the shapes of $P(\RMHost|star\ density)$ and $P(\RMIGM|primordial)$ are almost identical, as their peaks move through the same value at this redshift.
This causes the two contributions to greatly match each other, resulting in an identical shape of the full $P(\RMEG)$.
In contrast to that, $\RMIGM$ values in the \astrophysical scenario are to weak to significantly alter the shape of $P(\RMHost)$.
Hence, the two IGM scenarios appear very similar at that redshift.
This cosmic conspiracy might be used to infer the strength of the primordial magnetic field $B$, as the position of the dip highly depends on $B$.
However, it also strongly depends on the shape of the contribution of the host galaxy and might not be visible for other sets of models. \\

\new{We've shown that there likely exists at least a subset of FRBs - produced by magnetars in wind environments hosted by starburst dwarf galaxies - that carries information on the IGMF.
    However, other host galaxies and progenitor classes completely overshadow that signal of the IGM.
    This shows how important it is to carefully consider the numerous possible models for regions along the LoS and to identify the host galaxyies and source objects in order to measure the IGMF using FRBs.
    }

According to Bayes' theorem, in order to arrive at the ratio of posterior likelihoods $L$  for the different models, the Bayes factor $B$ has to be multiplied by the prior corroboration $\pi(M)$ towards a model $M$, based on information other than the investigated observation $O$ \citep[e. g.][]{imagine_whitepaper,stat4astroBayes}:
\begin{equation}
    \frac{L(M_1|O)}{L(M_2|P)} 
    =
    b(O|M_1,M_2) \times \frac{ \pi(M_1)}{\pi(M_2)}
\end{equation}
Note that we assume the two IGM models to be equally likely, i. e. $\pi = const$.
It therefore suffices to investigate the Bayes factor $b$ to infer the posterior likelihood of different models.

\section{Discussion}
\label{sec:discussion}
\new{We investigate whether observations of fast radio bursts can deliver information about the intergalactic magnetic field and its origins.}
To this end, we consider two extreme scenarios of magneto-genesis: a scenario where the intergalactic magnetic field is seeded at very high redshift (termed \primordial scenario) and a second scenario where the magnetic field is mainly supplied by galactic outflows and other astrophysical processes (termed \astrophysical scenario).

The initial magnetic field is very different in the two scenarios.
Hence, the two scenarios account for a strong difference in the strength of magnetic fields far outside the overdense regions in the Universe.
This implies significantly different results for the rotation measure \new{and makes these two suitable models to investigate the potential of fast radio bursts to probe the intergalactic magnetic field.}
We compute likelihood functions of these measures that allow a comparison of the assumed models to observational data. \\

To account for the contribution towards the rotation and dispersion measures of the host, we investigated two models for the host galaxy, i. e. a MW-like spiral galaxy and a starburst dwarf galaxy similar to IC~10 or the host of FRB121102. This only serves as an illustration of our framework to compare theory and observations and this framework can easily be expanded to include a large variety of models for the host galaxy.
Results agree with previous works \citep{Xu2015,Basu2018}.

Likewise, we model in Sec. \ref{sec:progenitor} the contribution of the local environment of the progenitor with Monte-Carlo simulations using the results of \citet{Piro2018}. 
The source of FRBs is assumed to be a magnetar embedded either in a uniform ISM or an environment disturbed by stellar winds of the seed star.
For the \uniform case, the distribution of possible ISM number densities $\nISM$ is determined by the host galaxy.
\\

In accordance with previous work \citep{dolag2015,Niino2018,Walker2018,Pol2019}, we find that the dispersion measure is an imprecise measure of the source redshift and only delivers reasonable upper limits.
Only few of the investigated models had low probability to supply $\text{DM}$ in excess of $\DMIGM$.
Hence, a significant fraction $\gtrsim 5$ per cent of high $\DMobs \gtrsim 10^3 \unitDM$ would point to a population of FRBs at cosmic distance, $z \gtrsim 1$.
However, this requires a more detailed investigation that takes into account the uncertain evolution of the number of FRB sources with redshift, as well as the selection effects of the telescopes.
We note that this is the scope of {\sc FRBpoppy}!\footnote{\href{https://github.com/davidgardenier/frbpoppy}{https://github.com/davidgardenier/frbpoppy}}, the results of which will be implemented in this framework in the future.\\

\citet{bhandari2017} report three FRBs with very high $\DMobs > 1500 \unitDM$ detected by the Parkes telescope.
Although the $\text{DM}$ are highly debated to be produced locally, they raise hope that there is indeed a FRB population at large distance that will be detected in the years to come.
For example, the MeerKat \& Parkes telescopes can detect FRBs out to redshift $z \approx 4$ \citep{Keane2018}. 
\new{
ARECIBO may detect bursts at $z \approx 5$ \citep{Lorimer2018}, while FAST will be able to detect FRBs even out to $z\approx 15$ \citep{Zhang2018}, with $\DMobs$ exceeding $10^4 \unitDM$ .
Our study shows that these FRBs are an interesting source of information on the IGMF and its origins.
}\\

For the limited set of models investigated in this paper, only a few progenitor models are capable to produce the high $\text{RM}$ observed for FRB121102 \citep{Michilli2018}.
For other FRBs with less extreme $\text{RM}$s, conclusions on their source are less obvious and require careful investigation of the convolved likelihood functions of the different contributors.
The time evolution of repeating FRBs can be used to put much better constraints on the source model.
This is, however, beyond the scope of this study and will be considered in upcoming work. \\

The models applied for the host galaxy use analytic functions and do not account for local over-densities, that can add significantly to $\RMHost$. 
Also, our models of the host do not yet account for cosmic evolution of the galaxy. 
Results of \citet{pillepich2019} suggest that low-mass star-forming galaxies do not change their size significantly out to redshift $z=4-5$. 
Hence the values of $\DMHost$ are not expected to change much for the dwarf-type of galaxies considered here. 
They further find that massive galaxies similar to the Milky Way tend to be smaller at higher redshift. 
The density can be higher by a factor of few tens, while the path length is reduced by a factor of a few, accounting for a $\text{DM}$ higher by about 1 order of magnitude than predicted in our model, still mostly below the contribution of the IGM.

For magnetic fields in galaxies, the amplification time is of order $10^7 - 10^8 \rm\ yr$ \citep[e. g.][]{schober2013}.
Observations and simulations of galaxies at high redshift suggest magnetic fields of similar strength as in present-day galaxies \citep{Kronberg2008,Bernet2008,pakmor2014,Mao2017}.
Hence, the expected change to the $\RMHost$ is of the same order as for the $\DMHost$, insignificant for dwarf-type and about one order of magnitude higher for galaxies similar to the Milky Way.
This implies that the latter type of host dominates the extragalactic contribution and does not allow for conclusions on the IGMF, even out to redshift $z=6$.
However, \citet{rodrigues2018} conclude that a significant fraction of massive spiral galaxies contain negligible large-scale magnetic fields at redshifts $z > 3$, suggesting a significantly weaker host contribution.
A more physical modelling of the host galaxies will be the subject of future work.\\ 

We account for effects from the possible progenitor positions by testing different distributions within the host galaxy.
We find that assuming a uniform distribution in the host disc affects the distribution of expected $\DMHost$ only at values $\lesssim \rm\ few \unitDM$, as compared to a distribution that concentrates on the centre of the galaxy.

The distribution of expected $\RMHost$ is very different, even $\gtrsim 10^1 \unitRM$, close to the maximum possible value, with a much higher probability in the centered case because the highest $\RMHost$ come from the centre of the galaxy.
We note, however, that our model does not include the high $\text{RM} \approx -5.6 \cdot 10^5 \unitRM$ found for Sagittarius A$^\star$ \citep{marrone2006}.
Such contributions from a LoS through the galactic centre of the host galaxy might explain the high $\text{RM}$ observed for FRB121102 \citep{Michilli2018}.
We argue that such LoS are very unlikely for progenitors that are not themselves located in the centers of their host galaxy.\\

By assuming only magnetars as progenitors, we restrict the parameter space for equations in \citet{Piro2018}, as compared to neutron stars with weaker magnetic fields, e. g. Pulsars.
By that, we mostly exclude lower values of $\RMProg$, hence arrive at rather optimistic predictions for the contribution from the local environment of the progenitor. 
\new{Further, their model assumes that a coherent magnetic field is produced in shocks of super nova remnantsshocks in super nova remnants produce a coherent magnetic field, while it most likely has a random topology.
    Hence, the results for the \uniform and the \windSNR model should be considered as upper limits.
    In case the real contribution of RM from such magnetars is significantly lower, these sources might also deliver information on the IGMF.
}

We do not account for the contributions of the MW halo.
For the $\text{DM}$, they are comparable to the contribution from the Galactic disc, $\approx 30 - 80 \unitDM$ \citep{dolag2015,Prochaska2019}.
The $\text{RM}$ from the halo is probably lower than from the disc, due to the weaker magnetic fields.
However, the likelihood function of the two models for the IGM show reasonable difference at redshift $z=0.5$ even for $\RMEG > 1 \unitRM$.
We only used $\RMEG$ above this value in estimates of the model likelihood in Sec. \ref{sec:fake_test}. \\

We do not account for the distribution of galaxies that host FRBs.
By applying a constant weight to each LoS, we implicitly assume a uniform distribution of host galaxies.
Reducing the weight of LoS through low density regions mostly reduce the likelihood of low values of RM that cannot be probed by telescopes.
Estimating the effect on likelihood of RM $\gtrsim 1 \unitRM$ is not trivial and will be studied in upcoming works.
\new{However, for FRBs beyond redshift $z\gtrsim 0.1$, the overall statistics are not expected to change, since the universe is homogeneous on large scales. } \\

We do not account for the contribution of intervening galaxies \citep[e. g. ][]{Basu2018}.
If the intervening galaxy is of the same type as the host, the contribution is comparable to the host contribution at the redshift of intersection and therefore hampers the investigation of the IGM component \citep{zheng2014}.
If the pulse broadening of FRB radiation is found to be dominated by scattering in intervening galaxies, this can help to exclude LoS with a significant contribution of intervening galaxies \citep{Lorimer2013,spitler2014}.
This will be a subject of future studies. \\

Our results are provided in the form of likelihood functions for the different contributions to RM and DM.
We show how these likelihood functions can be used for parameter inference. 
This framework can help to infer the origin of FRBs as well as the origin of extragalactic magnetic fields \citep{caleb2018,palaniswamy2018,Katz2018}.\\

\section{Conclusions}
\label{sec:conclusion}
In this paper we have studied the different contributions to the dispersion measures and rotation measures along the lines-of-sight of Fast Radio Bursts.
We have built a Bayesian framework to interpret observable information of FRBs.
We show how this can be used to constrain the amplitude of magnetic fields in the IGM along the line-of-sight.
Our key findings are:
\begin{itemize}
        
    \item the strengths of the different contributions to the observed $\RMobs$ highly depend on the assumed model for FRBs and their host galaxies. 
    Magnetars embedded in wind environments hosted by starburst dwarf galaxies provide the lowest average local contribution to RM of the investigated models.
    For this generous set of models, $\RMEG$ from redshifts $z \gtrsim 0.5$ can potentially provide information on the magnetic field in the IGM and its origin.
    At this redshift the contribution of the intergalactic medium is still subdominant to that of the host.
    Still, there is a significant change in the distribution of extragalactic $\RMEG$.
    This allows one to draw conclusions on the magnetic field using Bayesian inference.

    Conversely, for other models of the host galaxy and progenitor, the expected local contribution can be significantly stronger.
    These models require FRBs beyond $z \gtrsim 3$  in order to probe IGMFs.
    \new{We conclude that there are good reasons to believe that at least a subset of FRBs observed with RMs deliver information on the IGMF and its origin.
    How to identify this subset will be subject of future studies.}
    
    \item the Milky Way provides the dominant contribution of RM, even for FRBs out to redshift $z \gtrsim 6$.
    A prerequisite for the result above is the removal of the contribution of the MW to a precision of $\sim 1 \unitRM$.
    This is non trivial, as argued by \citet{2017ARA&A..55..111H}, who suggested that values up to $\text{RM} \sim 10^4-10^5 \unitRM$ may be necessary to tell apart Galactic from extragalactic contributions. 
    However, the fast growing level of complexity in modelling magnetic fields in the MW is expected to improve at the same pace as $\text{RM}$ statistics, making the removal of the MW foreground more viable \citep[for a recent review see e.g.][]{imagine_whitepaper}. 

    \item using likelihood functions allows one to infer information on the host galaxy and progenitor.
    They allow to systematically rule out models for a single FRB or groups of those.
    
    \item from the present set of models, only some progenitor models are capable of producing the very high $\RMobs$ of FRB121102.
    Our results suggest that, if the progenitor is a magnetar, then it is most likely located in a dense environment, such as an H\,{\sc ii}-region, \new{which we found to be capable of producing RM that exceed those of FRB121102 by two orders of magnitude}.
    \new{Note, however, that the strong magnetic fields generated by shocks in the super nova remnants are likely random.
        This can result in much lower RM than predicted by the model of \citet{Piro2018}, who assumed a coherent magnetic field.}
    
    We find a much smaller chance that stellar winds of the seed star in other environments can account for the high $\RMobs$ as well.
    The shape of magnetic fields induced by stellar winds is very coherent and can account for very high values of RM.
    However, the expected $\RMProg$ falls rapidly with age of the magnetar.
    This implies a much lower chance to observe high $\RMProg$ from such a source.

\end{itemize}

We provide a framework for the comparison between observations and theories of fast radio bursts.
So far, we consider a very limited set of models in order to present our framework.
\new{Still, we could show the likely existence of a subset of fast radio bursts that delivers information on the intergalactic magnetic field and its origins.}
Future work will include more models, such as elliptical or disc host galaxies, and take into account their evolution with redshift.
\new{We will vary the strength of a primordial magnetic field in realistic combination with astrophysical processes.
    This will allow to precisely probe the average strength of intergalactic magnetic fields today as well as the strength of the primordial seed field, thus allow to constrain processes of magneto-genesis with fast radio bursts.
    }

At this point, we only consider the dispersion measure and rotation measure.
In future work, more observables will be considered, such as temporal scattering, flux density and fluence.
Further contributing regions will be considered, such as intervening galaxies and the halo of the Milky Way. 
Combining this with knowledge on the selection effect of telescopes and assumptions on the underlying population of FRBs, we can produce individual predictions for the distribution of observables as measured at different telescopes.\footnote{This is the aim of the {\sc PreFRBLE} (\say{Predictions of Fast Radio Burst models and their Likelihood Estimates}) python package that comes with the presented results for a fast application to observational data.
This package is currently under construction and will be publicly available soon.}

\section*{Acknowledgements}
We thank Tony Piro for his help in understanding and reproducing results from his paper.
We also thank Ryan Mckinven for interesting comments.
Further, we thank the anonymous referee for a detailed review and constructive comments.
S. H. would like to thank all participants of \say{A Bayesian View on the Galactic Magnetic Field} as well as the Lorentz Center for hospitality during the program.
S. H. would also like to thank the organizers, speakers and participants of the \say{Stat4Astro School of Statistics for Astrophysics 2017: Bayesian Methodology} and the \say{FRB2019} conference in Amsterdam for interesting discussions.

Our cosmological simulations were performed with the {\sc ENZO} code (http://enzo-project.org), under project HHH38 and HHH42 at the J\"ulich Supercomputing Centre (P.I. F. Vazza). 
F. V. acknowledges financial support from the ERC  Starting Grant \say{MAGCOW}, no. 714196.
We thank Jenny Source and Stefan Gottl\"ober for providing us with the CLUES initial conditions and for their help in implementation. 

The Dunlap Institute is funded through an endowment established by the David Dunlap family and the University of Toronto. B. M. G.. acknowledges the support of the Natural Sciences and Engineering Research Council of Canada (NSERC) through grant RGPIN-2015-05948, and of the Canada Research Chairs program.




\bibliographystyle{mnras}
\bibliography{paper} 




\appendix
\section{Uniform primordial magnetic field}
\label{app:uniform}
\begin{figure}
    \centering
    \includegraphics[width=0.47\textwidth]{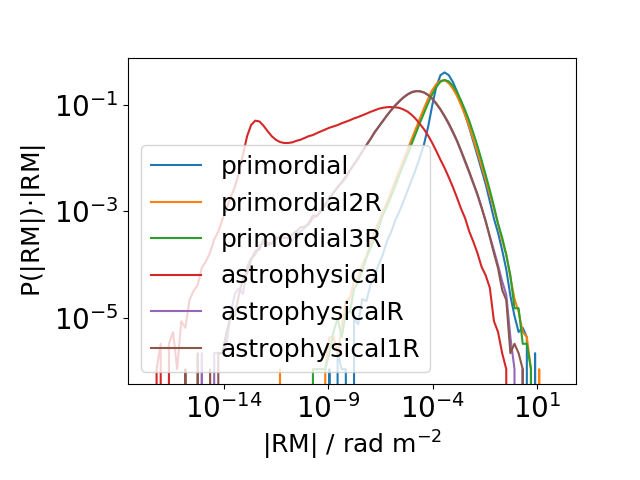}
    \caption{likelihood function $P(\text{RM})$ for the different models of the local Universe from \citet{Hackstein2018}.
    }
    \label{fig:uniform}
\end{figure}
{
In this section we briefly investigate the effects of choosing a uniform primordial magnetic field on the resulting $\text{RM}$ likelihood functions.
To this end, we compare the six models presented by \citet{Hackstein2018}.
In this paper we used their \primordial model, which starts with a uniform magnetic field.
The other \primordial models start from a purely turbulent field using different power law indices, while the \astrophysical models use a very faint seed field and instead allow for mangetic feedback from AGN.  \\
Using the uniform resolution grid at $z=0$, we calculate $\text{RM}$ for LoS parallel to two axes, positive and negative direction, to obtain both signs for RM values.
From that we obtain likelihood function of $\text{RM}$ for the different configuration of initial magnetic field.

The result in Fig. \ref{fig:uniform} shows that for the \primordial model, which starts from a completely uniform magnetic field, there is a pronounced peak at around $10^{-3} \unitRM$.
The other {\it primordial2R} and {\it primordial3R}  models start from a stochastic field, so contributions along the LoS can cancel out each other, the likelihood reduces at the peak value and increases at lower values.
Higher values, contributed by denser stuctures, are not affected by the shape of the primordial field.
However, this feature is visible in the IGM component at order $\RMIGM \approx 10^{-4} \unitRM$, and hence not accessible.
This implies that $\text{RM}$s of FRBs do not carry information on the coherence length of primordial fields.  \\

The \astrophysical models show similar highest values of $\RMIGM$ to the \primordial ones.
The bulk of values is a few orders of magnitude below the \primordial peak and the low tail reaches to substantially smaller values.
The shape at low values is rather different from results in Fig. \ref{fig:likelihood_RM_IGM} at redshift $z=1$, as it reaches to smaller values and peaks again at around $10^{-13} \unitRM$.
The differenece is because predictions in this work have been reconstructed from the \primordial model.
However, since the difference in results is for values of $\text{RM}$ that are far too low to be measurable, we consider the data sufficient for the argument of this work.
}
\section{Priors}
\label{sec:priors}
\newpage
\begin{table*}
\begin{tabular}{l|l|l}
\multicolumn{3}{l}{\bf Priors:} \\ \hline \hline
\bf Host Galaxy & \\ \hline \hline
 \multirow{2}{*}{position of progenitor $pos$} &  MW: $\prod\limits_{i \in [\rm thin, thick] }  e^{-\frac{z}{Z_i}} e^{-\frac{r}{R_i}}$ 
 & \citet{siegel2002,juric2008} \\ 
  & IC10: $e^{-\frac{z}{Z}} e^{-\frac{r}{R}}$ & \citet{leroy2006IC10} \\ \hline
  magnetic field of host $B_{\rm host}$ & IC10: log-flat, $B_{\rm host} \in [5 \cdot 10^{-1},5] \rm\ \mu G $ 
  & \citet{chyzy2016} \\
\hline \hline
\bf Progenitor & \\ \hline \hline
magnetic field of magnetar $B_{\rm NS}$ & LogNorm($log(2.5\cdot 10^{14}\rm G)$, 0.5) 
 & \citet{Ferrario2008} \\ \hline 
mass of SN ejecta $M$ & $M = m^{-2.35} - M_{\rm NS}$, $m \in \{20,45\} M_\odot$ 
  & \citet{chabrier2003} \\ \hline
\multirow{2}{*}{ISM number density $n_{\rm ISM}$ } & $ MW: \sum \frac{p_i}{\Delta n} \left( \Theta( n - n_i) - \Theta( n_{i+1} - n )  \right) $
   & \citet{Ferriere2001} \\ 
 & IC10: log-flat, $n_{\rm ISM} \in [5\cdot 10^{-3},3] $ 
 & \citet{Avillez2005} \\ \hline
time since SN $t$ & flat, $t \in \{ 25 yr, t_{\rm diss} \}$ & \citet{margalit2018, beloborodov2016}  \\ \hline
wind mass loading parameter $K$ & log-flat, $K \in \{ 10^{11},  10^{15} 
\} \rm g\ cm^{-1}$
 & \\ \hline
 magnetic field of seed star $B_\star$  & log-flat, $B_\star \in \{ 800, 1500 \} \rm G$  & \citet{Ferrario2008} \\ \hline

\end{tabular}
\caption{ Parameters for Monte-Carlo simulations, their prior distributions together with a reference }
\label{tab:priors}
\end{table*}

\begin{figure}[t]
    \includegraphics[width=0.48\textwidth]{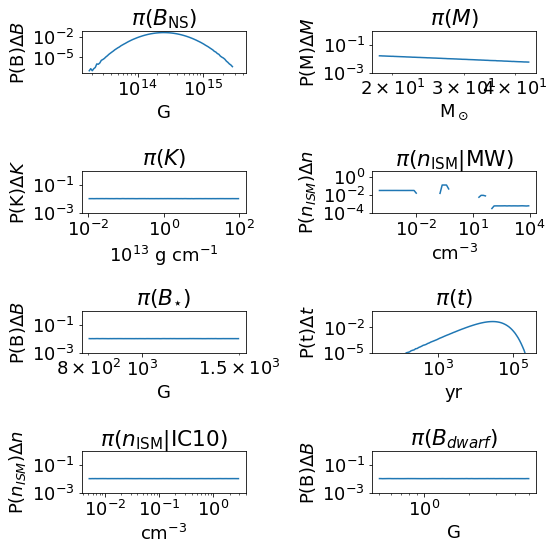}
    \centering
    \caption{Graphical depiction of the priors in the Tab \ref{tab:priors}.}
\end{figure}

We perform Monte-Carlo simulations in order to obtain likelihood functions for the contribution of the progenitor and the host galaxy.
This requires a choice of reasonable prior probability distribution of the parameters that enter the equations.
All parameters and their priors are summarized in Tab. \ref{tab:priors}. \\

We assume the source of FRBs to be magnetars, which stem from B- and O-type stars with masses of $m = 20-45 \rm M_\odot$ and $B_\star = 800 - 1500$ G dipole magnetic fields.
This assumption is reasonable, according to the results of \citet{Ferrario2008}.
The $\text{DM}$ and $\text{RM}$ contribution for this case of the local environment is given by \citet{Piro2018}. \\

\citet{Woosley1995} showed, that such supernovae explode with a typical energy of $E = 1.2 \cdot 10^{51} \rm erg$, which we adopt as a constant value. \\

The mass of neutron stars is about $ M_{\rm NS} \approx 1.5 M_\odot$, regardless of the progenitor stars mass.
Hence, the mass of supernova ejecta is the mass of the progenitor star $m$ minus the mass of the neutron star.
The prior of the mass of the progenitor star is given by the initial mass function, well approximated by the Salpeter function $\pi(M) \propto M^{-2.35}$ \citep{salpeter1955,chabrier2003,chabrier2005}, and has a support in the mass range stated above, reduced by the mass of the neutron star.

We obtain the stellar radius from the radius-mass relation of heavy stars given in \citet{demircan1990}. \\

The number density of the ISM, $n_{\rm ISM}$, in a MW-like galaxy is highly varied across the different media found throughout the galaxy.
We use the ranges of $n_{\rm ISM}$ given by \citet{Ferriere2001} together with the well known volume filling factors of the different media.
Within each of these ranges, we choose a log-flat distribution, renormalized, such that the integral over the range gives the volume filling factor of the corresponding medium.

For IC10, we assume a constant $n_{\rm ISM}$ throughout the disk of the dwarf galaxy that falls exponentially with scale height.
Since FRBs are mostly located in the disc, we use identical priors for the progenitor and the host galaxy. \\

In general, the production of FRBs is not related to the supernova that gives birth to the magnetar.
Hence, no age of the magnetar is preferred over another, which is reflected by a flat prior.
Free-free absorption by supernova ejecta can weaken FRB radiation.
This implies a lower limit of $t \gtrsim 25 \rm\ yr$ on the age of magnetars to emit visible FRBs \citep{margalit2018}.
We adopt this value as a strict lower limit.
The activity period of magnetars is limited by the dissipation of their strong magnetic field, $10^{14} - 10^{16}$ G.
The dissipation timescale was derived by \citet{beloborodov2016}:
\begin{equation}
	t_{\rm diss} = 600 \ufrac{L}{ 1 km }^{1.6} \ufrac{\delta B_{\rm NS}}{10^{16} G}^{0.4} \ufrac{B_{\rm NS}}{10^{16} G}^{-1.6} \ufrac{ \rho }{ \rho_{\rm nuc} }^{1.2} \rm yr,
\end{equation}
where $L$ is the typical scale of variation, $\delta B_{\rm NS}$, of the magnetar's magnetic field strength, $B_{\rm NS}$. $\rho$ is the density of the magnetar and $\rho_{\rm nuc} = 2.8 \cdot 10^{14} g\ cm^{-3}$ is the nuclear saturation density.
While the magnetic field dissipates, FRBs become less likely.
We account for that by sampling possible values of $t_{\rm diss}$ and use the shape above the maximum of the resulting probability density function.

The dissipation time, $t_{\rm diss}$, depends on parameters that are independent of all other parameters of interest.
To sample $t_{\rm diss}$, we assume typical values of $L = 10^5 \rm\ cm$  and $\rho = 10^{14} \rm g\ cm^{-3}$ \citep{beloborodov2016}.
For the magnetic field of the magnetar, $B_{\rm NS}$, we roughly fit the results of \citet{Ferrario2008} with a LogNorm function centered at $2.5\cdot 10^{14} \rm G$. \\

The wind mass loading parameter, $K$ \citep{Piro2018}, is not well constrained so far.
Hence, we choose a log-flat prior in the expected range $K = 10^{11} - 10^{15} \rm g\ cm^{-1}$. \\

The distribution of strong magnetic fields in B- and O-type stars, $B_\star$, is rather uncertain, as is the relation between the magnetic field of the progenitor star and that of the magnetar.
This is because the strong field of the magnetar could stem from either a fossil field or a shear driven dynamo.
As a conservative choice, we use a log-flat prior for $B_\star$ and consider $B_\star$ and the magnetic field of the magnetar, $B_{\rm NS}$ as independent.

By definition, $B_{\rm NS} = 10^{14} - 10^{16} \rm\ G$.
\citet{Ferrario2008} give the distribution of $B_{\rm NS}$ for observed magnetars and their best fit model.
In order to minimize selection effects from observations, we adopt their best fit model as a LogNorm with mean $\mu = 2.5\cdot 10^{14}\rm\ G$ and logarithmic deviation $\sigma = 0.5$

For simplicity, we assume that $\delta B_{\rm NS} \sim B_{\rm NS}$, which is generally the case for magnetic fields of such strength \citep[see e. g. ][]{beloborodov2016}. \\

For the host galaxy we investigate two different models: a MW-like spiral galaxy and a dwarf galaxy similar to IC10. \\

We assume that the probability for the position of an FRB scales with the number density of stars.
In the MW-like galaxy, the best fit model is the combination of two disks, thin and thick, with exponential decay from centre towards the borders.
We use the best fit parameters for the MW, given in \citet{juric2008}, $Z_{\rm thin}=0.3 \rm\ kpc$, $R_{\rm thin}=2.6 \rm\ kpc$, $Z_{\rm thick}=0.9 \rm\ kpc$, $R_{\rm thick} =3.6 \rm\ kpc$ \\

The distribution of stars in dwarf galaxies like IC10 is irregular.
We hence use a simple disk model with scale height $Z = 300 kpc$ \citep{leroy2006IC10} and radius $R = 900 kpc$. \\

\citet{Avillez2005} provide a distribution of ISM density, $n_{\rm ISM}$, in star forming galaxies that is well described by a log-flat distribution and can be used as prior distribution of $n_{\rm ISM}$ in dwarf galaxies like IC10. \\

\citet{chyzy2016} give a range of possible strengths for the ordered magnetic field of the dwarf galaxy IC10.
We do not assume a particular shape, as the number of values is too low to derive a reasonable distribution.
Hence, we use a log-flat distribution that covers the range of these values. \\

At this point, we take all our models as candidates with equal prior likelihood.
The ratio of their inferred posteriors is hence equal to the ratio of their measure likelihoods, i. e. the Bayes factor.




\label{lastpage}
\end{document}